\documentclass[11pt,a4paper]{article}
\usepackage{hyperref}
\usepackage{epsfig}
\usepackage{amsmath}
\usepackage{amssymb}
\usepackage{color}
\usepackage[footnotesize]{caption}
\usepackage[subrefformat=parens,labelformat=parens]{subfig}
\usepackage{cite}

\numberwithin{equation}{section}

{
{

\begin{document}

\begin{center}
{\Large\bf  $A_4$ realization of left-right symmetric linear seesaw}
\\[2mm]
\vskip 2cm

{Purushottam Sahu$^{a}$\footnote{purushottams@iitbhilai.ac.in},
Sudhanwa Patra$^{a}$ \footnote{sudhanwa@iitbhilai.ac.in},
Prativa Pritimita$^{b}$\footnote{prativa@iitb.ac.in}}\\[3mm]
{\it{
$^a$Dept. of Physics, Indian Institute of Technology Bhilai, Raipur-492015, India,\\ $^b$ Dept. of Physics, Indian Institute of Technology Bombay, Powai, Mumbai-400076, India}
}
\end{center}

\vskip 1cm

\begin{abstract}

\noindent 
We explore an $A_4$-symmetric flavor based left-right symmetric model with linear seesaw mechanism and study the associated neutrino phenomenology. The framework offers the advantage of studying neutrino mass, non-unitarity effects in lepton sector, lepton flavour violation and CP violation.  The fermion content of the model includes usual quarks, leptons along with additional sterile fermion per generation while the scalar content includes Higgs doublets and scalar bidoublet. 
We study analytically as well as numerically the correlation between different model parameters and their dependence on experimentally determined neutrino observables. 

\end{abstract}

\newpage

\section{Introduction}\label{sec1}

The left-right symmetric model (LRSM) which was initially proposed as 
the most economical approach to restore parity came a long way in explaining neutrino mass, lepton number violation, 
dark matter, baryon asymmetry of the universe thereby gaining popularity~\cite{Mohapatra:1974gc, Pati:1974yy, Senjanovic:1975rk, Senjanovic:1978ev,
Mohapatra:1980yp,Akhmedov:1995vm,Heeck:2015qra,Garcia-Cely:2015quu,Patra:2015vmp,Patra:2015qny}. All accolades to its gauge group, 
i.e. $SU(3)_C \times SU(2)_L \times SU(2)_R \times U(1)_{B-L}$ which naturally gives room to a right handed neutrino 
and obeys a complete symmetry between left and right chirality untill spontaneous symmetry breaking 
occurs. The model predicts 
$W_R^{\pm}$ and $Z^\prime$ gauge bosons which couple to Standard Model fields and heavy Majorana neutrino $N$ 
and these exotic states at low scale offer rich collider phenomenology. Neutrino mass 
can be explained within LRSM via various seesaw mechanisms like the canonical seesaw \cite{Minkowski:1977sc,Mohapatra:1979ia,Dev:2013oxa,Agostinho:2017wfs,Magg:1980ut,Schechter:1980gr,Cheng:1980qt,Lazarides:1980nt} and its lowscale variants like inverse seesaw~\cite{Dev:2009aw}, linear seesaw, 
extended seesaw etc~\cite{Mohapatra:2005wg}. The issue with canonical seesaw is that it links the smallness of neutrino mass to a very heavy right-handed scale which can't be 
verified by the current or planned experiments. Whereas in case of low scale seesaw heavy neutrino mass can lie in the TeV range which is 
experiment friendly and moreover it offers rich phenomenology like lepton flavor violation~\cite{Cirigliano:2004mv,Cirigliano:2004tc,Riazuddin:1981hz,Pal:1983bf,Mohapatra:1992uu} and new physics contributions to lepton number violating decays like 
neutrinoless double beta decay~\cite{Deppisch:2017vne, Hati:2018tge, Pritimita:2016fgr,Ge:2015bfa, Deppisch:2014zta, Tello:2010am,Deppisch:2014qpa,Patra:2014goa,Borah:2013lva,Awasthi:2013ff,Patra:2012ur,Chakrabortty:2012mh,Dev:2013vxa,Barry:2013xxa,Nemevsek:2011hz,Dev:2014iva,Keung:1983uu,Das:2012ii,Bertolini:2014sua,Beall:1981ze,Ge:2015yqa,Deppisch:2015cua,Hirsch:1996qw,Dev:2013oxa,Dev:2015pga,Bajc:2009ft}. 
Inverse seesaw and linear seesaw can be realized with the introduction of extra singlet fermions per generation 
to the LRSM where the mass matrix in the basis $(\nu_L, ~\nu^c_R, ~S_L)$ can be written as,
\begin{equation}
	\mathbb{M} = 
	\begin{pmatrix}
		0 & m_\text{D} & m_\text{LS} \\
		m_\text{D}^\text{T} & 0 & m_{RS} \\
		m_\text{LS}^\text{T} & m^\text{T}_{RS} & \mu
	\end{pmatrix}.		
\end{equation}
Thus the light neutrino mass formula becomes, 
$m^{\rm inv}_\nu = \left(\frac{m_D}{m_{RS}} \right) \mu \left(\frac{m_D}{m_{RS}} \right)^T$ in the former case and 
$m^{\rm lin}_\nu = m_\text{D} m_{RS}^{-1} m_\text{LS}^\text{T} \mbox{+ transpose}$ in the later case. It can be 
interpreted from the formula that it allows 
order one magnitude of Dirac Yukawa coupling, large light-heavy neutrino mixing and heavy neutrinos of mass few TeV. 
In \cite{Pritimita:2016fgr}, LRSM has been extended to study natural 
type-II seesaw dominance which allows large light-heavy neutrino mixing thus leading to many new physics contributions 
to neutrinoless double beta decay along with constraints on light neutrino mass. Another interesting variant is 
universal seesaw  which allows all the quarks and leptons to get mass from a common seesaw due to the addition of 
vector-like fermions in LRSM \cite{Deppisch:2017vne}.

However we aim to study here the $A_4$ extension of LRSM which offers the advantage of studying neutrino mass, non-unitarity effect in linear seesaw, 
lepton flavour violation and CP violation within one framework. $A_4$, the discrete group of even permutations of four objects is the smallest non-Abelian group 
containing triplet irreducible representations. While it was first proposed in \cite{Ma:2001dn} to study lepton masses and mixing, 
several other $A_4$-based flavour models have been suggested after that mostly to shed light on the 
flavour problem \cite{Babu:2002dz,Altarelli:2005yp,Holthausen:2012wz,King:2013hj}. In one such recent work \cite{Karmakar:2016cvb} the authors 
have elegantly explained the origin of non-zero $\theta_{13}$ 
and leptogenesis via inverse seesaw with $A_4$ extension of SM. However in the model the light-heavy neutrino 
mixing is proportional to identity and thus the branching ratios of 
LFV decays become vanishingly small. Similarly, another paper which considers realization of linear seesaw with $A_4$ 
symmetry gives suppressed contributions to lepton flavour violating (LFV) decays due to the chosen model parameters \cite{Sruthilaya:2017mzt}. This can be ameliorated by 
considering $A_4$ realization of LRSM where large non-unitarity effect can be achieved and thus it can lead to 
dominant contributions to LFV decays, which is the primary motive of this work.
The embedding of $A_4$ group into left-right flavour symmetry has been attempted previously in order to explain charged fermion mass hierarchies 
and quark and lepton mixing angles \cite{Morisi:2007ft,Ma:2006wm}.

In this work, we have considered $A_4$ realization of left-right symmetric linear seesaw mechanism. The fermion 
content of the model includes usual quarks, leptons along with additional sterile fermion 
per generation while the scalar content includes Higgs doublets with $B-L=1$ and scalar bidoublet 
with $B-L=0$. Within this scenario all the fermion masses get simpler mass structure for neutrino phenomenology. 
While the right-handed Higgs doublet $H_R$ plays the crucial role of left-right symmetry breaking, its left handed partner $H_L$ has merely any role. 
Thus the non-zero VEV of $H_L$ can be taken too small. As usual the scalar bidoublet $\Phi$ plays the role of electroweak 
symmetry breaking. The introduction of $A_4$ symmetry helps to avoid the $\mu$ term and hence, inverse seesaw term 
for light neutrino masses is absent. The other good points of the model are large light-heavy neutrino mixing, prominent non-unitarity effects, 
dominant lepton flavour violating effects and CP-violating effects. The work also contains a number of plots showing 
correlation among model parameters and the experimentally observed parameters. 

The paper is organised as follows. In Sec \ref{sec:LRSM} we briefly note the features of manifest left-right symmetric model and then move to explain 
the realization of linear seesaw structure with $A_4$ extension of the left-right model. In Sec \ref{sec:mass-mixing} we discuss neutrino masses and mixing. We do so by 
setting up the flavour structure of neutrino mass matrices and establish analytically the correlation among model parameters. In Sec \ref{sec:numerical result} we estimate 
numerically the correlation among model parameters by using the values of experimentally measued neutrino parameters. Sec \ref{sec:non unitarity} explains non-unitarity effects 
in linear seesaw and Sec \ref{sec:LFV} has a discussion on various low energy lepton flavour violating processes that the model facilitates. In Sec \ref{sec:cp} we study 
leptonic CP violation for active neutrinos using Jarlskog invariants and in Sec \ref{sec:conclusion} we summerize the work.

\section{Left-right symmetric model with linear sessaw}
\label{sec:LRSM}
The gauge group of left-right symmetric model (LRSM) is given by 
\begin{equation}
 SU(3)_C \times SU(2)_L \times SU(2)_R \times U(1)_{B-L}
\end{equation}
where $B-L$ stands for the difference between baryon number and lepton number. The standard lepton and quark content of the model 
is given by 

\begin{eqnarray}
 \ell_L=\left(
\begin{array}{c}
\nu_{{L}}\\
e_{L}
\end{array}
\right)\sim (1,2,1,-1), \hspace{1cm}
\ell_R=\left(
\begin{array}{c}
\nu_{{R}}\\
e_{R}
\end{array}
\right)\sim (1,1,2,-1)\\
 q_L=\left(
\begin{array}{c}
u_{{L}}\\
d_{L}
\end{array}
\right)\sim (3,2,1,\frac{1}{3}), \hspace{1cm}
q_R=\left(
\begin{array}{c}
u_{{R}}\\
d_{R}
\end{array}
\right)\sim (3,1,2,\frac{1}{3}).
\end{eqnarray}

The scalar sector of a general LRSM contains two Higgs doublets and a bidoublet 
\begin{eqnarray}
 H_L=\left(
\begin{array}{c}
h^0_{L}\\
h^-_{L}
\end{array}
\right)\sim (1,2,1,-1), \hspace{1cm}
H_R=\left(
\begin{array}{c}
h^0_{R}\\
h^-_{R}
\end{array}
\right)\sim (1,1,2,-1),
\end{eqnarray}

\begin{eqnarray}
 \Phi=\left(
 \begin{array}{cc}
 \phi^0_1 & \phi^+_2\\
 \phi^{-}_1 & \phi^{0}_2
 \end{array}
\right)\sim (1,2,2,0). 
\end{eqnarray}

In order to generate neutrino mass through linear seesaw mechanism within LRSM, we will have to add one left-right gauge singlet
neutral fermion $S_L$ to the model. Now the complete Lagrangian for the leptonic sector becomes,

\begin{eqnarray}
 -\mathcal{L}_{\rm lepton}&=& 
\overline{\ell_L}\left(Y \Phi+\tilde{Y} \tilde{\Phi}\right)\ell_R
                        \nonumber\\
                         & &+Y_L \overline{\ell_L} H_L S_L + Y_R \overline{\ell_R} H_R S_L
                         +h.c.
\end{eqnarray}
Once the scalars $H_L, H_R$ and $\Phi$ obtain VEV as,
\begin{eqnarray}
\langle H_L \rangle=\left(
\begin{array}{c}
0\\
v_L
\end{array}
\right), \hspace{.5cm}
\langle H_R \rangle=\left(
\begin{array}{c}
0\\
v_R
\end{array}
\right), \hspace{0.5cm}
\langle \Phi \rangle=\left(
 \begin{array}{cc}
 v_1 & 0\\
 0 & v_2
 \end{array}
\right),  
\end{eqnarray}
the above Lagrangian can be written as,
\begin{eqnarray}
 -\mathcal{L}_{\rm lepton}&=& m_{LR} \overline{\nu_L}\nu_R
                              + m_{RS} \overline{\nu_R^c}S_L+ m_{LS} \overline{\nu_L} S_L + h.c.
\end{eqnarray}
Hence in the the basis  $(\nu_L, \nu_R^c, S_L)$, the effective $9\times 9$ mass 
matrix 
can be written as 
\begin{eqnarray}
M_{\nu} = \left(
\begin{array}{ccc}
         0     & m_{LR}     &m_{LS}\\
         m^T_{LR} & 0       &m_{RS}\\
         m^T_{LS}     & m^T_{RS}     & 0
\end{array}
\right),
\end{eqnarray}
where $m_{LR}=Y_1 v_1 + Y_2 v_2^*$ 
and $m_{RS}=y_1 v_R$. 
In the above matrix, one may wonder why the $\mu$ term is absent, which will be clarified in the next section once we introduce $A_4$ symmetry.
Now, with $m_{LS} << m_{LR} < m_{RS}$, the light neutrino masses are obtained from the formula,
\begin{eqnarray}\label{mnu1}
 m_{\nu}&=&m_{LR} m^{-1}_{RS} m^T_{LS} + \mbox{transpose} 
\end{eqnarray}

\subsection{An $A_4$ realization of left-right symmetric linear seesaw mechanism}

\begin{table}
\begin{center}
\begin{tabular} {|c|c|c|c|c|c|c|c|c|}\hline
& Fields & $SU(3)_c$ & $SU(2)_L$ & $SU(2)_R$ & $B-L$ & $A_4$ & $Z_4$ & $Z_3$\\ 
\hline
& $\ell_{L_{1,2,3}}$ & {\bf 1} & {\bf 2} & {\bf 1} & $-1$ & $1,1'',1'$ & $-1$ &1\\
& $\ell_{R}$ & {\bf 1} & {\bf 1} & {\bf 2} & $-1$ & $3$ & $i$ 
&1\\
& $S_L$ & {\bf 1} & {\bf 1} & {\bf 1} & 0 & 3 & $i$ &$\omega^2$ \\
\hline
& $\Phi$ & {\bf 1} & {\bf 2} & {\bf 2} & 0 & 1  &1 &1\\
& $H_L$ & {\bf 1} & {\bf 2} & {\bf 1} & $-1$ & 1 & 1 & $\omega$\\
& $H_R$ & {\bf 1} & {\bf 1} & {\bf 2} & $-1$ & 1 & 1 &$\omega^2$\\
\hline 
& $\phi_{S}$ & {\bf 1} & {\bf 1} & {\bf 1} & $0$ & 3 & $-1$ &1\\
& $\phi_{T}$ & {\bf 1} & {\bf 1} & {\bf 1} &  $0$ & 3 & $i$ &1\\
& $\xi$ & {\bf 1} & {\bf 1} & {\bf 1} & $0$ & 1 & $-1$ &1\\
& $\xi'$ & {\bf 1} & {\bf 1} & {\bf 1} &  $0$ & $1'$  & $-1$ &1\\
\hline
\end{tabular}
\end{center}
\caption{Particle content and transformation properties under
$A_4$ based flavour left-right symmetric model. }
\label{tab:1}
\end{table}

$A_4$ symmetry group has four irreducible representations; three singlets, namely $1, 1^\prime, 1^{\prime\prime}$ and a triplet $3$. 
The multiplication rules for the irreducible representations can be written as; 
$1\otimes1=1$, $1^\prime\otimes1^\prime=1^{\prime\prime}$, $1^{\prime\prime}\otimes1^{\prime\prime}=1^\prime$, $1^\prime\otimes1^{\prime\prime}=1$. 
The product of two $A_4$ triplets; ($a_1,a_2,a_3$) and ($b_1,b_2,b_3$) 
can be written as,
\begin{eqnarray}
3\times 3&=&1+1^{\prime}+1^{\prime\prime}+3_A+3_S
\end{eqnarray}
\begin{eqnarray}
1\sim a_1b_1+a_2b_3+a_3b_2 \\ \nonumber
1^{\prime}\sim a_3b_3+a_1b_2+a_2b_2 \\ \nonumber
1^{\prime\prime}\sim a_2b_2+a_3b_1+a_1b_3 \\ \nonumber
3_S \sim \frac{1}{3}\left[\begin{array}{c}
2a_1b_1-a_2b_3-a_3b_2\\
2a_3b_3-a_1b_2-a_2b_1\\
2a_2b_2-a_1b_3-a_3b_1
\end{array} \right]\\ \nonumber
3_A\sim \frac{1}{2}\left[\begin{array}{c}
a_2b_3-a_3b_2\\
a_1b_2-a_2b_1\\
a_3b_1-a_1b_3
\end{array} \right]\\ \nonumber
\end{eqnarray}
With the particle content and symmetries mentioned in Table \ref{tab:1}, the 
Lagrangian involved in generation of the mass matrices in a left-right $A_4$ flavor symmetric framework can be written as, 
\begin{align}\label{a4lag}
 -\mathcal{L}_{\rm lepton}=&\mathcal{L}_{\nu_L \nu_R}+\mathcal{L}_{\nu_R S_L}+\mathcal{L}_{\nu_L S_L}\,,
\end{align}
where
\begin{align}
 \mathcal{L}_{\nu_L \nu_R}= \label{lag:diracmass}
&\frac{1}{\Lambda}(\overline{\ell_{L_1}})_{1} \left(Y \Phi+\widetilde{Y} \tilde{\Phi}\right) \left(\ell_R \phi_T\right)_{1} \nonumber \\
&+\frac{1}{\Lambda} (\overline{\ell_{L_2}})_{1^{\prime \prime}} \left(Y \Phi+\widetilde{Y} \tilde{\Phi}\right) \left(\ell_R \phi_T\right)_{1^\prime} \nonumber \\
&+\frac{1}{\Lambda} (\overline{\ell_{L_3}})_{1^\prime} \left(Y \Phi+\widetilde{Y} \tilde{\Phi}\right) \left(\ell_R \phi_T\right)_{1^{\prime \prime}}  
\end{align}

With the vevs for the scalar and flavon fields as,
$\langle \phi_S \rangle= v_S(1,1,1), \langle \phi_T \rangle= v_T(1,0,0), \langle\xi \rangle=v_{\xi}, \langle\xi^{\prime}\rangle
=v_{\xi^{\prime}}$, 
we obtain  the flavor structures of the involved mass matrices. 

\begin{eqnarray}\label{mCL}
M_{\ell} =\frac{v_T}{\Lambda} \left(
\begin{array}{ccc}
         Y_{11} v_2+\widetilde{Y}_{11} v_1^* & 0                 &  0\\
         0                 & Y_{22} v_2+\widetilde{Y}_{22} v_1^* &  0\\
         0                 & 0                 & Y_{33} v_2+\widetilde{Y}_{33} v_1^*
\end{array}
\right).  
\end{eqnarray}
In analogy to the charged lepton mass matrix, the Dirac mass for light neutrinos can be written as, 
\begin{eqnarray}
m_{LR} &=&\frac{v_T}{\Lambda} \left(
\begin{array}{ccc}
         Y_{11} v_1+\widetilde{Y}_{11} v_2^* & 0                 &  0\\
         0                 & Y_{22} v_1+\widetilde{Y}_{2} v_2^* &  0\\
         0                 & 0                 &  Y_{33} v_1+\widetilde{Y}_{33} v_2^*
\end{array}
\right)\\
     &=&\frac{v_T}{\Lambda} Y_{D}v \label{md} \left(
\begin{array}{ccc}
         1 & 0  &  0\\
         0 & 1  &  0\\
         0 & 0  &  1
\end{array}
\right), 
\end{eqnarray}
where we have considered, $Y_D v \equiv Y v_1+\widetilde{Y} v_2^*=Y_{11,22,33} v_1+\widetilde{Y}_{11,22,33} v_2^*$. 

It is seen from Table-1 that under $A_4$, $\ell_L$ transforms as $1, 1^{\prime}, 1^{\prime \prime}$ for 1st, 2nd and 3rd generation of left-handed doublet 
respectively, $S_{L}$ transforms as triplet and $H_L$, $H_R$ as singlets. Thus the $\nu_L-S_L$ mixing term; $\overline{\ell^c_L} H_L S_L$  and 
the $\nu_R-S_L$ mixing term $\overline{\ell_R} H_R S_L$ are not allowed at tree level and are generated at dimension five level as follows,
\begin{align}
 \mathcal{L}_{\nu_L S_L} &= \frac{Y^{11}_L}{\Lambda} (\overline{\ell_{L_1}})_{1} 
   H_L (S_L \phi_T)_{1}  \nonumber \\
&+\frac{Y^{22}_L}{\Lambda} (\overline{\ell_{L_2}})_{1^{\prime \prime}} 
   H_L (S_L \phi_T)_{1^{\prime}} \nonumber \\
&+\frac{Y^{33}_L}{\Lambda} (\overline{\ell_{L_3}})_{1^{ \prime}} 
   H_L (S_L \phi_T)_{1^{\prime \prime}} 
\end{align}
Once $\langle H_L \rangle$ and $\langle \phi_T \rangle$ get vev, the $\nu_L-S_L$ mixing matrix becomes,
\begin{eqnarray}
m_{LS} &=&\frac{v_T}{\Lambda} \left(
\begin{array}{ccc}
         Y^{11}_L v_L & 0                 &  0\\
         0                 & Y^{22}_L v_L &  0\\
         0                 & 0                 &  Y^{33}_L v_L
\end{array}
\right)\\
     &=&\frac{v_T}{\Lambda} Y_{L} v_L \label{mLS} \left(
\begin{array}{ccc}
         1 & 0  &  0\\
         0 & 1  &  0\\
         0 & 0  &  1
\end{array}
\right), 
\end{eqnarray}
where we have considered, $Y_L v_L =Y^{11,22,33}_L v_L$. As we have mentioned earlier, the scalar $H_R$ plays the crucial role of 
left-right symmetry breaking, and $H_L$ is required only for left-right invariance. Thus $H_L$ gets a small induced vev which is much smaller than $H_R$. 
This clarifies why $m_{LS}$ term is much smaller than the $m_{RS}$ term.

Similarly, the $\nu_R-S_L$ mixing term is generated at dimension five level as follows

\begin{align} 
\mathcal{L}_{\nu_R S_L} & =  \frac{1}{\lambda}  \left(\lambda^{\phi_s} \phi_s 
                            +\lambda^{\xi} \xi+
                          \lambda^{\xi^{\prime}} \xi^{\prime} \right) \overline{\ell_R} H_R S_L\;.
\end{align}
The advantage of forbidding $\overline{\nu_L}-S_L$ and $\overline{\nu_R}-S_L$ terms at tree level and generating them by dimension five operator is that 
it helps in achieving large light-heavy neutrino mixing which gives large non-unitarity effects and lepton flavour violation. 
It should be noted that the terms $\mathcal{L}_{\nu_L \nu_R}$, $\mathcal{L}_{\nu_L S_L}$ and $\mathcal{L}_{\nu_R S_L}$ represent the contributions 
for Dirac neutrino mass connecting $\nu_L-\nu_R, \nu_L-S_L$, $\nu_R-S_L$ mixing, respectively. 
If one looks at the mass formula for light neutrinos governed by linear seesaw mechanism given in Eq.\ref{mnu1}, 
one can use the mass hierarchy $m_{RS} \gg m_{LR}, m_{LS}$. 

Using the   following vevs for the scalar and flavon fields
\begin{eqnarray}\label{e4a}
\langle \phi_S \rangle= v_S(1,1,1), \langle \phi_T \rangle= v_T(1,0,0), \langle\xi \rangle=v_{\xi},~ \langle\xi^{\prime}\rangle
=v_{\xi^{\prime}},~~
\end{eqnarray}
the various mass matrices are found to be,
\begin{eqnarray}
&&m_\text{LR}= \frac{Y_D v v_T}{\Lambda}\left(
\begin{array}{ccc}
1 & 0 & 0\\
0 & 1 &0 \\
0 & 0 &1
\end{array}
\right)
,~~~~m_\text{LS}=\frac{Y_L v_L v_T}{\Lambda}
\left(
\begin{array}{ccc}
1 & 0 & 0\\
0 & 1 &0 \\
0 & 0 &1
\end{array}
\right), \label{e4-b} \\ \label{e4-a}
&&m_{RS}=\frac{a}{3}\left(
\begin{array}{ccc}
2 & -1 & -1\\
-1 & 2 &-1 \\
-1 & -1 &2
\end{array}
\right)+
b\left(
\begin{array}{ccc}
1 & 0& 0\\
0 & 0 &1 \\
0& 1 &0
\end{array}
\right)+
d\left(
\begin{array}{ccc}
0 & 0& 1\\
0 & 1 &0 \\
1& 0 &0
\end{array}
\right),
\end{eqnarray}
where $a=\lambda^{\phi}v_S v_R/\Lambda$, $b=\lambda^{\xi}v_{\xi} v_R/\Lambda$ and $d=\lambda^{\xi^{\prime}} v_{\xi^{\prime}} v_R /\Lambda$.

\section{Neutrino Masses and Mixing}
\label{sec:mass-mixing}
In this section we focus on studying the correlation between different model parameters and their dependence on experimentally determined neutrino 
parameters. We start by rewriting the mass matrix $m_\text{RS}$ (\ref{e4-a}) for calculational convenience as,
\begin{equation}
m_{RS}=\left(
\begin{array}{ccc}
2a/3+b & -a/3& -a/3\\
-a/3 & 2a/3 &-a/3+b \\
-a/3& -a/3+b &2a/3
\end{array}
\right)+
\left(
\begin{array}{ccc}
0 & 0& d\\
0 & d &0 \\
d& 0 &0
\end{array}
\right).
\end{equation}
The importance of this matrix is that it dictates the flavour structure of the light neutrino mass matrix $m_\nu$ in the linear seesaw scenario. 
Moreover, the structure of this matrix is such that, it leads to lepton mixing consistent with 
neutrino oscillation data ~\cite{Capozzi:2017ipn,Aker:2019qfn}.

Using Eqns.(\ref{e4a}),(\ref{e4-b}) and (\ref{e4-a}),
 one can write the light neutrino mass matrix as,
\begin{eqnarray}
m_{\nu}&=&m_{LR} m_{RS}^{-1} m_{LS}^T+\text{transpose} \nonumber\\
&=&k_1k_2\left(\begin{array}{ccc}
1 & 0 & 0\\
0 & 1 &0 \\
0 & 0 &1
\end{array}\right) m_{RS}^{-1}\left(\begin{array}{ccc}
1 & 0 & 0\\
0 & 1 &0 \\
0 & 0 & 1
\end{array}\right),
\end{eqnarray}
where the parameters $k_1$ and $k_2$  are related to the vevs through
\begin{eqnarray}
k_1=\sqrt{2} Y_D v \frac{v_T}{\Lambda}\;, ~~~~k_2=\sqrt{2} Y_L v_L \frac{v_T}{\Lambda}.\nonumber
\end{eqnarray}
 The inverse of light neutrino mass matrix becomes,
\begin{eqnarray}
m_{\nu}^{-1}
&=&\frac{1}{k_1k_2}\left(
\begin{array}{ccc}
2a/3+b & -a/3& -a/3\\
-a/3 & 2a/3 &-a/3+b \\
-a/3& -a/3+b &2a/3
\end{array}
\right)+\frac{1}{k_1k_2}\left(
\begin{array}{ccc}
0 & d& 0\\
d & 0 &0 \\
0& 0 &d
\end{array}
\right),
\end{eqnarray}
which in TBM~\cite{Harrison:2002er,Harrison:1999cf,Rashed:2011zs} basis  will have the form, i.e., $m_{\nu}^{-1^{\prime}}=U_\text{TBM}^T m_{\nu}^{-1}U_\text{TBM}$, 
\begin{equation}
m_{\nu}^{-1^{\prime}}=\left(\begin{array}{ccc}
a+b-d/2 & 0 & -\frac{\sqrt{3}}{2}d\\
0 &~~ b+d~~ &0 \\
-\frac{\sqrt{3}}{2}d & 0 &a-b+d/2
\end{array}\right)\;.
\end{equation}
The above matrix $m_{\nu}^{-1^{\prime}}$ can be  diagonalized by $U_{13}^*$. Which means $m_\nu^{-1}$  
can be diagonalized by $U_{TBM}\cdot U_{13}^*$ and $m_{\nu}$ by $U_{TBM}\cdot U_{13}$, while $m_{RS}$  by $U_{TBM}\cdot U_{13}^T$. 
The matrix $U_{13}$ and $U_{TBM}$ are of the form 
\begin{equation}
 U_{TBM}=\left(
 \begin{array}{ccc}
 \sqrt{\frac{2}{3}} & \sqrt{\frac{1}{3}} & 0 \\
 -\sqrt{\frac{1}{6}} & \sqrt{\frac{1}{3}} & \sqrt{\frac{1}{2}}\\
 -\sqrt{\frac{1}{6}}& \sqrt{\frac{1}{3}} & -\sqrt{\frac{1}{2}}
 \end{array}
 \right),\label{eqnTBM}
 \end{equation}
\begin{equation}
 U_{13}=\left(
 \begin{array}{ccc}
 \cos\theta & 0 & \sin\theta e^{-i \delta}\\
 0 & 1 & 0 \\
 -\sin\theta e^{i\delta}& 0 & \cos\theta
 \end{array}
 \right),\label{eqn14}
 \end{equation}
 where internal mixing angle $\theta$ and phase $\delta$ are expressed in terms of the 
 mass matrix parameters  $d/b=\lambda_1e^{\phi_{db}}$, $a/b=\lambda_2e^{\phi_{ab}}$ as
 \begin{eqnarray}
 \tan 2\theta=-\frac{\sqrt{3}\lambda_1\cos\phi_{db}}{(\lambda_1 \cos\phi_{db}-2)\cos\delta +(2\lambda_2\sin\phi_{ab})\sin\delta}\;,\label{tant}
 \end{eqnarray}
 and
 \begin{eqnarray}
 \tan\delta=\frac{\sin\phi_{db}}{\lambda_2\cos(\phi_{ab}-\phi_{db})}\;.\label{tanpsi}
 \end{eqnarray}
 The purpose of rotating the $m_\nu$ matrix by $U_{TBM}$ followed by $U_{13}$ is to achieve non-zero reactor mixing angle $\theta_{13}$ and see the possible 
 correlations between various parameters.
 
 Again, from Eq.3.1 and 3.2 it is found that eigenvalues of $m_{\nu}$ and $m_{RS}$ are related to each other as
 \begin{equation}
 m_i=\frac{k_1k_2}{M_i}\;.\label{e16}
 \end{equation}
  where $m_i$ and $M_i$ are $i^\text{th}$ eigenvalues of $m_{\nu}$ and $m_{RS}$ respectively. The eigenvalues of $m_{RS}$  can be expressed using 
  defined parameters in terms of $\lambda_1$ and $\lambda_2$ as,
  \begin{eqnarray}
M_1&=&b\left[\lambda_2 e^{i\phi_{ab}}-\sqrt{1+\lambda_1^2e^{2i\phi_{db}}-\lambda_1 e^{i\phi_{db}}}\right], \nonumber\\
M_2&=&b\left[1+\lambda_1 e^{i\phi_{db}}\right], \nonumber\\
M_3&=&b\left[\lambda_2 e^{i\phi_{ab}}+\sqrt{1+\lambda_1^2e^{2i\phi_{db}}-\lambda_1 e^{i\phi_{db}}}\right],\label{eq17a} 
 \end{eqnarray}

After some simple calculations, one can  write the heavy neutrino masses as
 \begin{eqnarray}
 M_1&=& |b|\left[(\lambda_2\cos\phi_{ab}-C)^2+(\lambda_2\sin\phi_{ab}-D)^2\right]^{1/2} ,\nonumber\\
 M_2&=& |b|\left[1+\lambda_1^2+2\lambda_1\cos\phi_{db}\right]^{1/2}, \nonumber\\
 M_3&=& |b|\left[(\lambda_2\cos\phi_{ab}+C)^2+(\lambda_2\sin\phi_{ab}+D)^2\right]^{1/2}\;,\label{eq17} 
 \end{eqnarray}
 where
 \begin{eqnarray}
 C&=&\left[ \frac{A+\sqrt{A^2+B^2}}{2}\right]^{1/2}\;,~~~~~~~
 D=\left[ \frac{-A+\sqrt{A^2+B^2}}{2}\right]^{1/2}\;,\nonumber\\
 A&=&1+\lambda_1^2\cos 2\phi_{db}-\lambda_1\cos\phi_{db}\;,~~~~
 B=\lambda_1^2\sin 2\phi_{db}-\lambda\sin\phi_{db}\;.\label{abcd}
 \end{eqnarray}
 and the phases ($\phi_i$'s) of $M_i$, i.e., $M_i=|M_i|e^{i \phi_i}$  as
 \begin{eqnarray}
 \phi_1&=&\tan^{-1}\left[\frac{\lambda_2\sin\phi_{ab}-D}{\lambda_2\cos\phi_{ab}-C}\right]\;,\nonumber \\
 \phi_2&=&\tan^{-1}\left[\frac{\lambda_1\sin\phi_{db}}{1+\lambda_1\cos\phi_{db}}\right]\;, \nonumber\\
  \phi_3&=&\tan^{-1}\left[\frac{\lambda_2\sin\phi_{ab}+D}{\lambda_2\cos\phi_{ab}+C}\right].
 \end{eqnarray}
 The matrix which diagonalizes active neutrino mass matrix, $U_{\nu}$ is given by
 \begin{eqnarray}
 U_{\nu}&=&U_{TBM}\cdot U_{13}\cdot P\;,
\end{eqnarray}  
with $P=\text{diag}(e^{-i\phi_1/2},e^{-i\phi_2/2},e^{-i\phi_3/2})$.
\\
 and the lepton mixing matrix, known as PMNS matrix is given by \cite{Pontecorvo:1957qd,Maki:1962mu} 
\begin{equation}
U_{\rm PMNS}=U_\ell^{\dagger}\cdot U_{\nu}\;,
\end{equation}
Here $U_\ell=\mathbb{I}$, which implies,
\begin{equation}
U_{\rm PMNS}=U_{\rm TBM}\cdot U_{13}\cdot P, \label{eqn15}
\end{equation} 
and this looks to be in good agreement with the experimental observations \cite{Chao:2011sp,M.:2014kca}. 
The PMNS matrix can be parametrized in terms of three mixing angles ($\theta_{13}$, $\theta_{23}$ and $\theta_{12}$) 
and three phases (one Dirac phase $\delta_{CP}$, and two Majorana phases $\rho$ and $\sigma$) as
\begin{equation}
U_\text{PMNS}=\left( \begin{array}{ccc} c^{}_{12} c^{}_{13} & s^{}_{12}
c^{}_{13} & s^{}_{13} e^{-i\delta_{CP}} \\ -s^{}_{12} c^{}_{23} -
c^{}_{12} s^{}_{13} s^{}_{23} e^{i\delta_{CP}} & c^{}_{12} c^{}_{23} -
s^{}_{12} s^{}_{13} s^{}_{23} e^{i\delta_{CP}} & c^{}_{13} s^{}_{23} \\
s^{}_{12} s^{}_{23} - c^{}_{12} s^{}_{13} c^{}_{23} e^{i\delta_{CP}} &
-c^{}_{12} s^{}_{23} - s^{}_{12} s^{}_{13} c^{}_{23} e^{i\delta_{CP}} &
c^{}_{13} c^{}_{23} \end{array} \right) P^{}_\nu \;,\label{pmns}
\end{equation}
where $c_{ij}=\cos\theta_{ij}$ and $s_{ij}=\sin\theta_{ij}$ and $P_{\nu}=\text{diag}(1,e^{i\rho/2},e^{i\sigma/2})$. 
From Eqns. (\ref{eqn15}) and (\ref{pmns}), one can find 
\begin{eqnarray}
&&\sin\theta =\sqrt{\frac{3}{2}}\sin\theta_{13}~, \nonumber \\
&&\sin\delta_\text{CP}=-\frac{\sin\delta}{\displaystyle{\sqrt{1-\frac{3(2-3\sin^2\theta_{13})}{(1-\sin^2\theta_{13})^2}\sin^2\theta_{13}\cos^2\delta}}}\approx -\sin\delta\;.
\end{eqnarray}
The advantage of expressing $\theta$ and $\delta$ in this manner is that they become related to the mixing observables $\sin^2\theta_{13}$ and $\delta_{CP}$ 
respectively. $\sin^2\theta_{13}$ is known more precisely than $\delta_\text{CP}$, and thus in our calculation we fix $\theta$ by taking the best fit value 
of $\sin^2\theta_{13}$ and consider all possible values of $\delta$ for which $\delta_\text{CP}$ falls within its $3\sigma$ experimental range.
 Even though the solar mixing angle lies slightly on the higher side of the observed central value in this case, 
 i.e., $\displaystyle{\sin^2 \theta_{12} = 1/\left (3-2\sin^2 \theta \right )}$, it is still within the $3 \sigma$ range of the  observed data. \\

\section{Numerical results}
\label{sec:numerical result}
 In the previous section we set up the flavour structure of neutrino mass matrices and analytically established correlation among model parameters by 
 fixing $\delta$, $\theta$ and other parameters like $\phi_{ab}$, $\phi_{db}$. In this section, we intend to estimate numerically the inter-relation among 
 the model parameters by using the measured values of the ratio of two mass squared differences, r and the different mixing angles, 
 $\theta_{12}$, $\theta_{13}$, $\theta_{23}$. From Eqns. (\ref{e16}) the light neutrino masses are found to be 
 \begin{equation}
  m_i=\frac{|k_1k_2|}{M_i}\;.\label{lmass}
\end{equation}
 Since the solar mass squared difference,$\Delta m^2_{21}$ and atmospheric mass squared difference, $|\Delta m^2_{32}|$ 
 are measured in neutrino oscillation experiments, we calculate the mass squared differences from 
 Eqn. (\ref{lmass}) as,
\begin{eqnarray}
\Delta m^2_{21}&=&\left|\frac{k_1k_2}{b}\right|^2\left(\frac{1}{{M_2}^2}-\frac{1}{{M_1}^2}\right)\;, \nonumber \\
 \left|\Delta m^2_{31}\right|&=&\left|\frac{k_1k_2}{b}\right|^2\left|\left(\frac{1}{{M_3}^2}-\frac{1}{{M_1}^2}\right)\right|\;. \label{mdiff}
\end{eqnarray}
In order to find the ratio of the two mass squared differences,(r), one may substituting the set of Eqns. (\ref{eq17}) in the above equations, so that
 \begin{eqnarray}
 r&=&\frac{\Delta m_{21}^2}{|\Delta m_{31}^2|}=\left[\frac{(\lambda_2\cos\phi_{ab}+C)^2+(\lambda_2\sin\phi_{ab}+D)^2}{1+\lambda_1^2+2\lambda_1\cos\phi_{db}}\right] \nonumber\\
 &\times &\left[\frac{(\lambda_2\cos\phi_{ab}-C)^2+(\lambda_2\sin\phi_{ab}-D)^2-\left(1+\lambda_1^2+2\lambda_1\cos\phi_{db}\right)}{4\lambda_2|C \cos\phi_{ab}+D\sin\phi_{ab}|}\right]\;.\label{ra}
 \end{eqnarray}
Now using Eqs. (\ref{tant}), (\ref{tanpsi}), (\ref{eq17}), (\ref{abcd}) and (\ref{ra}), and by fixing the 
parameters $\phi_{db}$, $\delta$ and $\theta$, one can find numerical values of $M_i$. Once $M_i$ are 
known $\left|\displaystyle{\frac{k_1k_2}{b}}\right|$ can be calculated from (\ref{mdiff}) as
\begin{equation}
\left|\frac{k_1k_2}{b}\right|=\sqrt{\frac{\Delta m^2_{21}}{\displaystyle{ \left (\frac{1}{M_2^{2}}-\frac{1}{M_1^{ 2}}\right )}}}
 =\sqrt{\left|\frac{\Delta m^2_{31}}{\left (\frac{1}{M_3^{2}}-\frac{1}{M_1^{ 2}}\right )}\right|}\;,
\end{equation}
which will also give the absolute value of light neutrino masses as all the quantities on the right hand side of (\ref{lmass}) are now known.

 We now rewrite the expression  $\tan \delta$ Eq(\ref{tanpsi})  in terms of $\phi_{db}$ as
\begin{equation}
\phi_{db}=0,\pi,~~~\text{for}~ \tan\delta=0\;,
\end{equation}
and 
\begin{equation}
\phi_{ab}=\phi_{db}+ \cos^{-1}\left(\frac{\sin\phi_{db}}{\lambda_2\tan\delta}\right),~~~\text{for}~~ \tan\delta\neq 0,
\end{equation}
 and consider the following cases to see the implications.\\

\subsection{Correlation between model parameters with $\tan\delta=0$}
The input model parameters for neutrino mass analysis are, 
$$\lambda_1, \lambda_2, \phi_{db}, \phi_{ab}, \delta$$
 
For simplification, we chose $\tan\delta = 0$ and $\phi_{db}$ will be taken either $0$ 
or $\pi$. \\
{\underline{\bf {Case-I- $\tan\delta = 0$,\, $\phi_{db}=0$}}};\\

From eq. (\ref{tant}), the expressions that relates $\lambda_1$ with internal mixing angle $\theta$ is,
 \begin{equation} 
 \lambda_1=\frac{2\tan2\theta}{\sqrt{3}+\tan2\theta}\;,
 \label{eq:l1}
 \end{equation}
 The ratio of the mass square differences $r$ (\ref{ra}), satisfies the relation 
 \begin{eqnarray}
 &&r=\left[\frac{\lambda_2^2+2\lambda_2C\cos\phi_{ab}+C^2}{(1+\lambda_1)^2}\right]\left[\frac{\lambda_2^2-2\lambda_2C\cos\phi_{ab}+C^2-(1+\lambda_1)^2}{4\lambda_2|C\cos\phi_{ab}|}\right], \label{rb} \nonumber\\
 &\mbox{where,}& C=\sqrt{\frac{1-\lambda_1+\lambda_1^2}{2}} \, .
 \end{eqnarray}
 \begin{figure}[t!]
	\centering
	\includegraphics[width=0.48\textwidth]{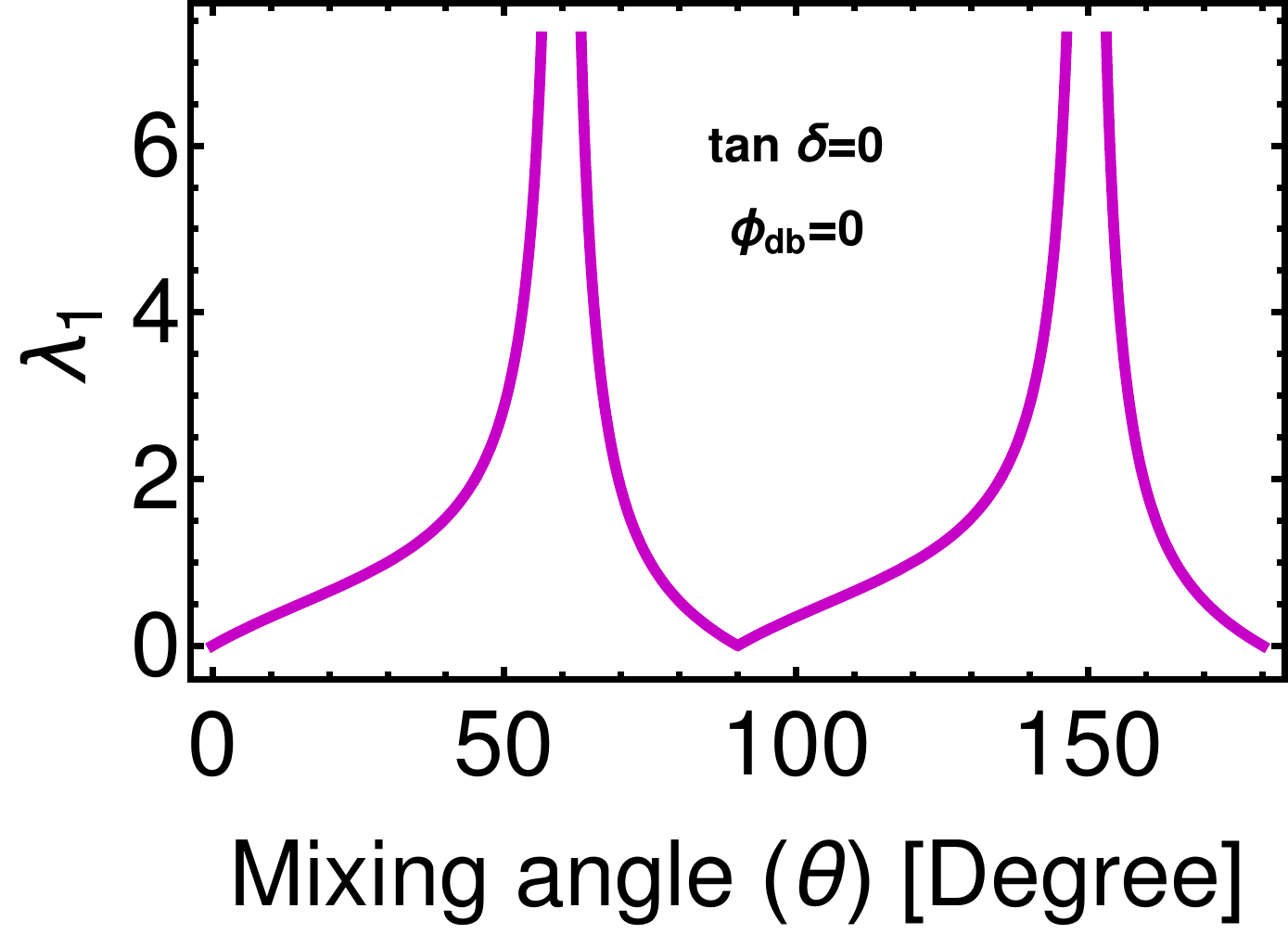}
		\includegraphics[width=0.48\textwidth]{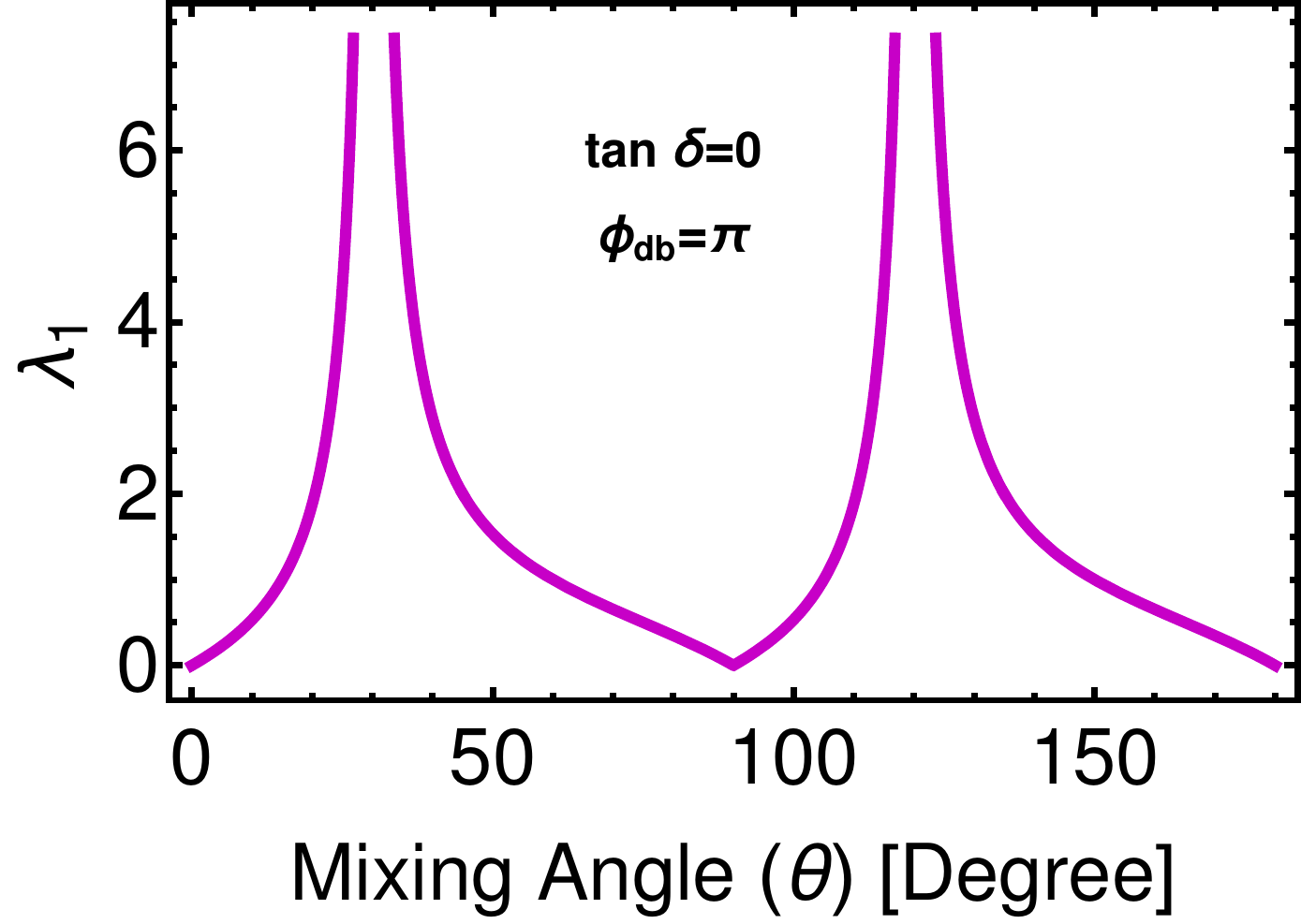}
	\caption{Plots for variation of $\lambda_1$ and internal mixing angle $\theta$ with $\tan{\delta}=0$ 
	and $\phi_{db}=0$ (left panel), $\phi_{db}=\pi$ (right panel).
	}
	\label{fig:l1-theta}
\end{figure}

\begin{figure}[t!]
	\centering
	\includegraphics[width=0.48\textwidth]{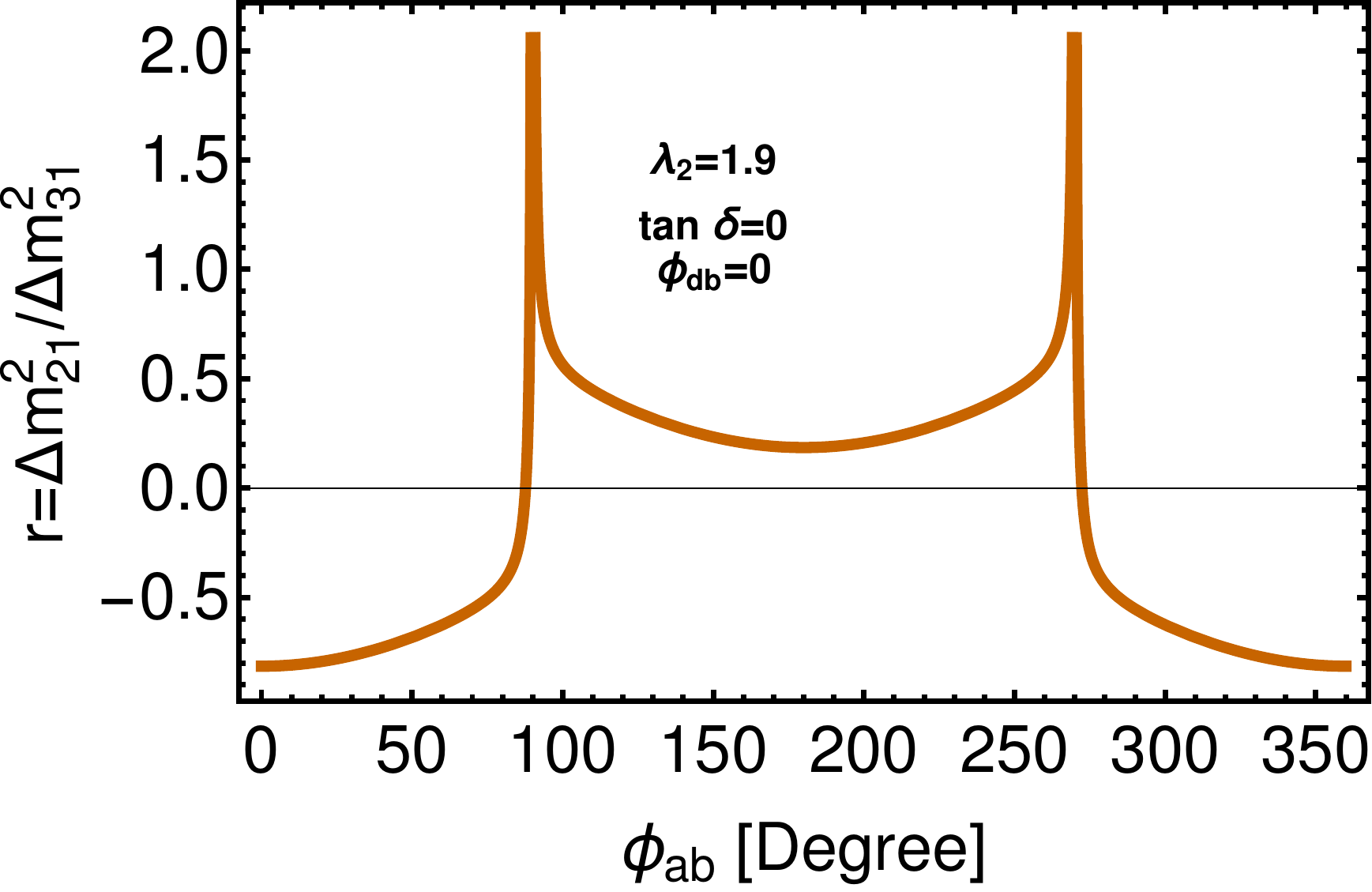}
		\includegraphics[width=0.48\textwidth]{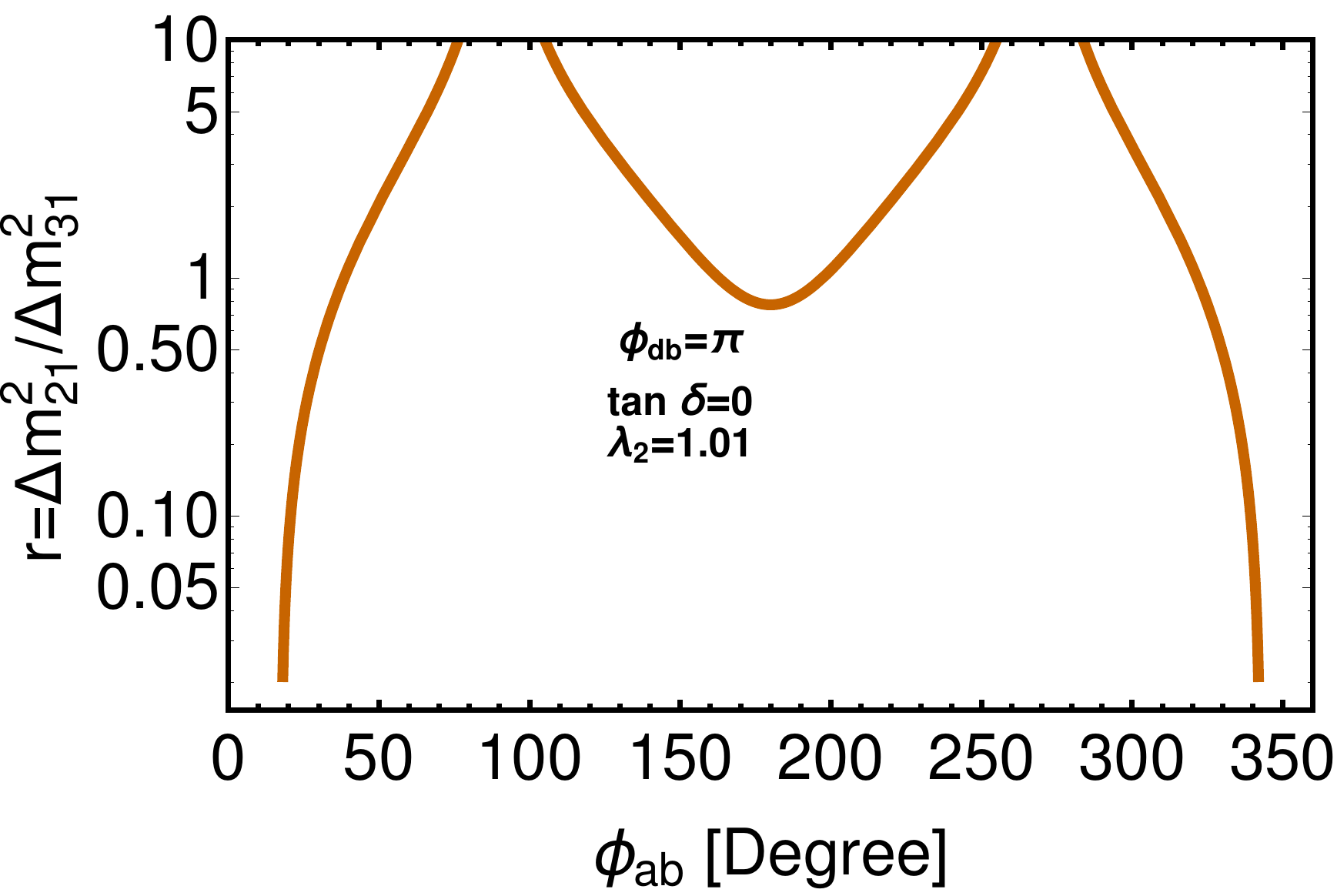}
	\caption{Contour plots for ratio of mass squared difference, r and $\phi_{ab}$ from $0$ to $\pi$ with $\phi_{db}=0$ (left panel) and $\phi_{db}=\pi$ (right panel)
	In these plots $\phi_{ab}$ is taken from $0$ to $2\pi$}
	\label{fig:Ratio-phiab2}
\end{figure}

The physical mass eigenvalues of $m_{RS}$ in this case become
 \begin{eqnarray}
 M_1&=&|b|\sqrt{\lambda_2^2-2\lambda_2C\cos\phi_{ab}+C^2}\;, \nonumber\\
 M_2&=&|b|(1+\lambda_1)\;,\nonumber\\  
 M_3&=&|b|\sqrt{\lambda_2^2+2\lambda_2C\cos\phi_{ab}+C^2}\;.
 \end{eqnarray}
Fig.\ref{fig:l1-theta} displays variation of input model parameter $\lambda_1$ with change in internal mixing angle $\theta$ in the range 0 to 180 degree.
It is seen from the figures that those values of $\theta$ which are multiples of $\frac{\pi}{4}$ are divergent or not allowed. The figures 
are plotted using equations (\ref{eq:l1}) and (\ref{eq:l11}) and it comes out that the figures are mirror images of each other due to the $`+ve'$ and $`-ve'$ signs of 
$\tan\,2\theta$ in the denomenators of the respective equations.
In Fig.\ref{fig:Ratio-phiab2} using eq.(\ref{rb}) it is shown that input model parameters with $\phi_{db}=0, \tan\delta=0$ and 
variation of phase angle $\phi_{ab}$ from $0-2\pi$ are consistent with experiment measured value of $r=0.03$~\cite{Capozzi:2017ipn}. 
In right-panel of Fig.\ref{fig:Ratio-phiab2}, it is shown that the ratio is divergent for $\phi_{ab}$ around 90 degree, which means $\phi_{ab}$ around 90 degree is not allowed. 
If we examine $\phi_{ab}$ from two different ranges, $0-\pi$ to $0-2\pi$ it is observed that for both case I and II the values 
of $\phi_{ab}$ around $(2n+1) \frac{\pi}{2} , n=1,2,3...$ are not allowed since at these values r diverges. So the constraints obtained on $\phi_{ab}$ is that $\phi_{ab} \neq (2n+1) \frac{\pi}{2} $.
 
 
{\underline{\bf{Case-II-$\tan\delta = 0$,\, $\phi_{db}=\pi$ }}} \\
For this case the relation between $\lambda_1$ and $\theta$ becomes,
\begin{equation} 
 \lambda_1=\frac{2\tan2\theta}{\sqrt{3}-\tan2\theta}\;,
 \label{eq:l11}
 \end{equation}
 and $r$ obeys the relation 
 \begin{equation}
 r=\left[\frac{\lambda_2^2+2\lambda_2C\cos\phi_{ab}+C^2}{(1-\lambda_1)^2}\right]\times\left[\frac{\lambda_2^2-2\lambda_2C\cos\phi_{ab}+C^2
 -(1-\lambda_1)^2}{4\lambda_2|C\cos\phi_{ab}|}\right], \label{r}
 \end{equation}
  with $C=\sqrt{\frac{1+\lambda_1+\lambda_1^2}{2}}$.
  
The eigenvalues of $m_{RS}$ can be written as 
 \begin{eqnarray}
 M_1&=&|b|\sqrt{\lambda_2^2-2\lambda_2C\cos\phi_{ab}+C^2}\;, \nonumber\\
 M_2&=&|b|(1-\lambda_1)\;,\nonumber\\  
 M_3&=&|b|\sqrt{\lambda_2^2+2\lambda_2C\cos\phi_{ab}+C^2}\;.
 \end{eqnarray}
For the above two cases we have shown the correlation plots in Fig.\ref{fig:l1-theta}. It should be noted from (\ref{rb}) that $r$ will be divergent near  $\phi_{ab}=\pi/2$ and thus, 
the values of $\phi_{ab}$ around $\pi/2$ are not allowed.

Similarly we can find Correlation between model parameters with $\tan\delta \neq 0$ . In this case The analytic expression for $\lambda_1$ is given by
 \begin{equation}
 \lambda_1=\frac{2\lambda_2\tan 2\theta\cos\phi_{ab}\sin\phi}{\sin\phi_{ab}\left[\sqrt{3}+\tan 2\theta\cos\phi\right]}\;.
 \end{equation}
 
\subsection{Analysis on Neutrino mixing angles}
%
 
In the present left-right symmetric model with linear seesaw mechanism, the light neutrino masses are diagonalized by $\mathbb{U}_{\rm TBM}$, $\mathbb{U}_{13}$ containing the 
 mixing angle $\theta$  and phases. The form of the mixing matrix  is expressed in terms of  $\theta$, $\delta$  and other phases in the 
 following way \cite{Sruthilaya:2017mzt, Karmakar:2016cvb,Mishra:2019oqq,Sethi:2019bfu},
 \begin{eqnarray}
 \mathbb{U} 
&&=\begin{pmatrix}
 \frac{2}{\sqrt{6}}  \cos\theta & \frac{1}{\sqrt{3}} & \frac{2}{\sqrt{6}} \sin\theta e^{-i \delta} \nonumber \\
 -\frac{1}{\sqrt{6}} \cos\theta+\frac{1}{\sqrt{2}}  \sin\theta e^{i\delta} & \frac{1}{\sqrt{3}} & -\frac{1}{\sqrt{6}} \sin\theta e^{-i\delta}-\frac{1}{\sqrt{2}}  \cos\theta \\
 -\frac{1}{\sqrt{6}}  \cos\theta-\frac{1}{\sqrt{2}}  \sin\theta e^{i\delta} & \frac{1}{\sqrt{3}} & -\frac{1}{\sqrt{6}}  \sin\theta e^{-i\delta}+\frac{1}{\sqrt{2}}  \cos\theta 
  \end{pmatrix}
  \cdot \begin{pmatrix}
 1 & 0 & 0\\
 0 & e^{\frac{i\alpha}{2}} & 0\\
 0 & 0 & e^{\frac{i\beta}{2}}
 \end{pmatrix}.
 \end{eqnarray}
 where $\alpha$ and $\beta$ are the two Majorana phases.
 
\begin{figure}[t!]
	\centering
	\includegraphics[width=0.65\textwidth]{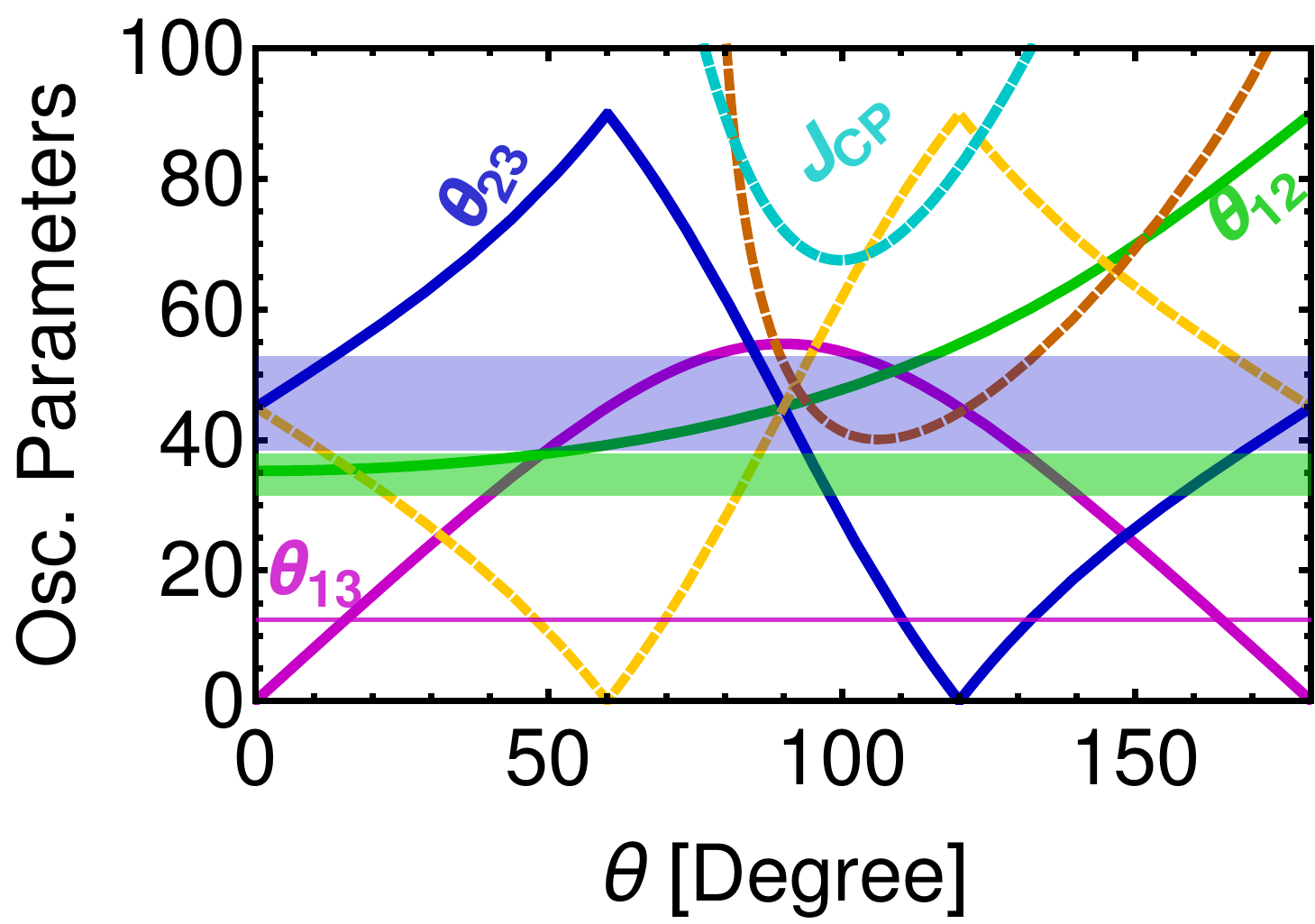}
	\caption{Variation of measured neutrino oscillation parameters like solar mixing angle ($\theta_{12}$), reactor mixing angle ($\theta_{13}$), and atmospheric mixing angle ($\theta_{23}$) with the change of internal mixing angle $\theta$ by fixing phase $\delta$.}
	\label{fig:mixingangles-theta}
\end{figure}
 
The neutrino mixing angles like solar mixing angle $\theta_{12}$, atmospheric mixing angle $\theta_{23}$, reactor mixing angle $\theta_{13}$ 
and Dirac CP-phase are related to the elements of the $U_ {\rm PMNS}$ through the following set of equations
$\sin^2 \theta_{13} = \mid \mathbb{U}_{e3} \mid^2$ , $\sin^2\theta_{12} =  \frac{\mid \mathbb{U}_{e2} \mid^2}{1 - \mid \mathbb{U}_{e3} \mid^2}$, 
$\tan^2 {\theta_{23}} = \frac{\mid \mathbb{U}_{\mu 3} \mid^2}{1 - \mid \mathbb{U}_{e3} \mid^2}$. The final expressions for these mixing angles can also be expressed in 
terms of the  model parameters like internal mixing angle $\theta$ and phase $\delta$ 
as,
\begin{eqnarray}
&& \sin^2 \theta_{13}  =\frac{2}{3}\sin^2\theta , \sin^2\theta_{12}  =\frac{1}{2+\cos2\theta}, \sin^2 {\theta_{23}} =\frac{1}{2}(1+\frac{\sqrt{3} \sin 2\theta \cos \delta}{2+\cos 2\theta}). \nonumber
 \end{eqnarray}
The other known quantiny in neutrino sector is Jarlskog rephrasing invariant~\cite{Jarlskog:1985cw} which can be expressed in terms $\theta$ and $\delta$ as,
 \begin{equation}
 J_{\rm CP}=\mbox{Im} \Big[\mathbb{U}_{e1} \mathbb{U}_{\mu 2} \mathbb{U}^*_{e2} \mathbb{U}^*_{\mu 1} \Big] =\frac{\sin\theta_{13}}{3\sqrt{2}}\sin\delta \sqrt{1-\frac{3}{2}\sin^2\theta_{13}}\, ,
 \end{equation}
 Using $\sin\theta_{13}\simeq 0.16$ and $|\sin \delta| > \frac{1}{2}$, the allowed range $0.026 < |J_{\rm CP} | < 0.036$ is obtained.

\begin{figure}[t!]
	\centering
	\includegraphics[width=0.45\textwidth]{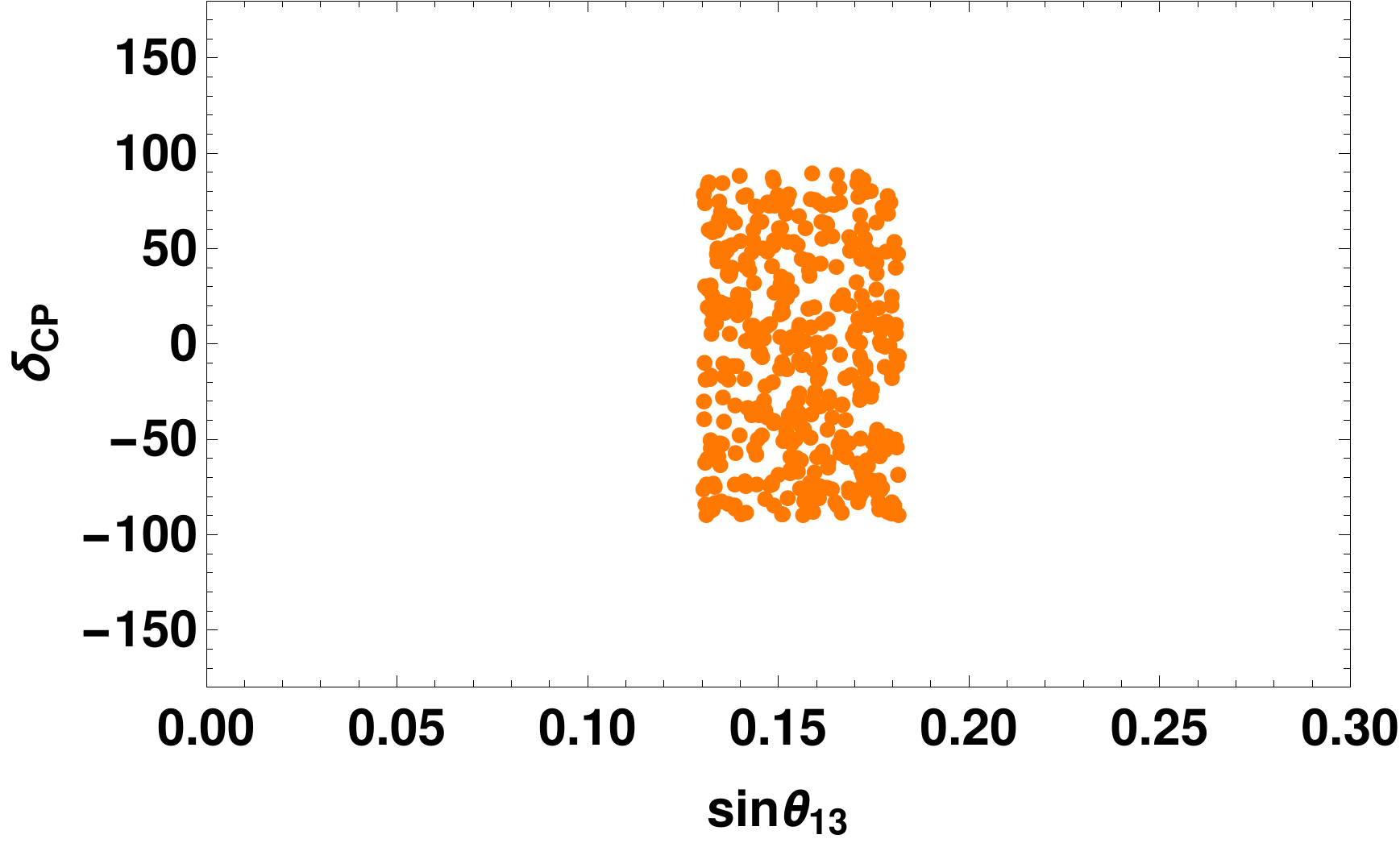}
	\includegraphics[width=0.45\textwidth]{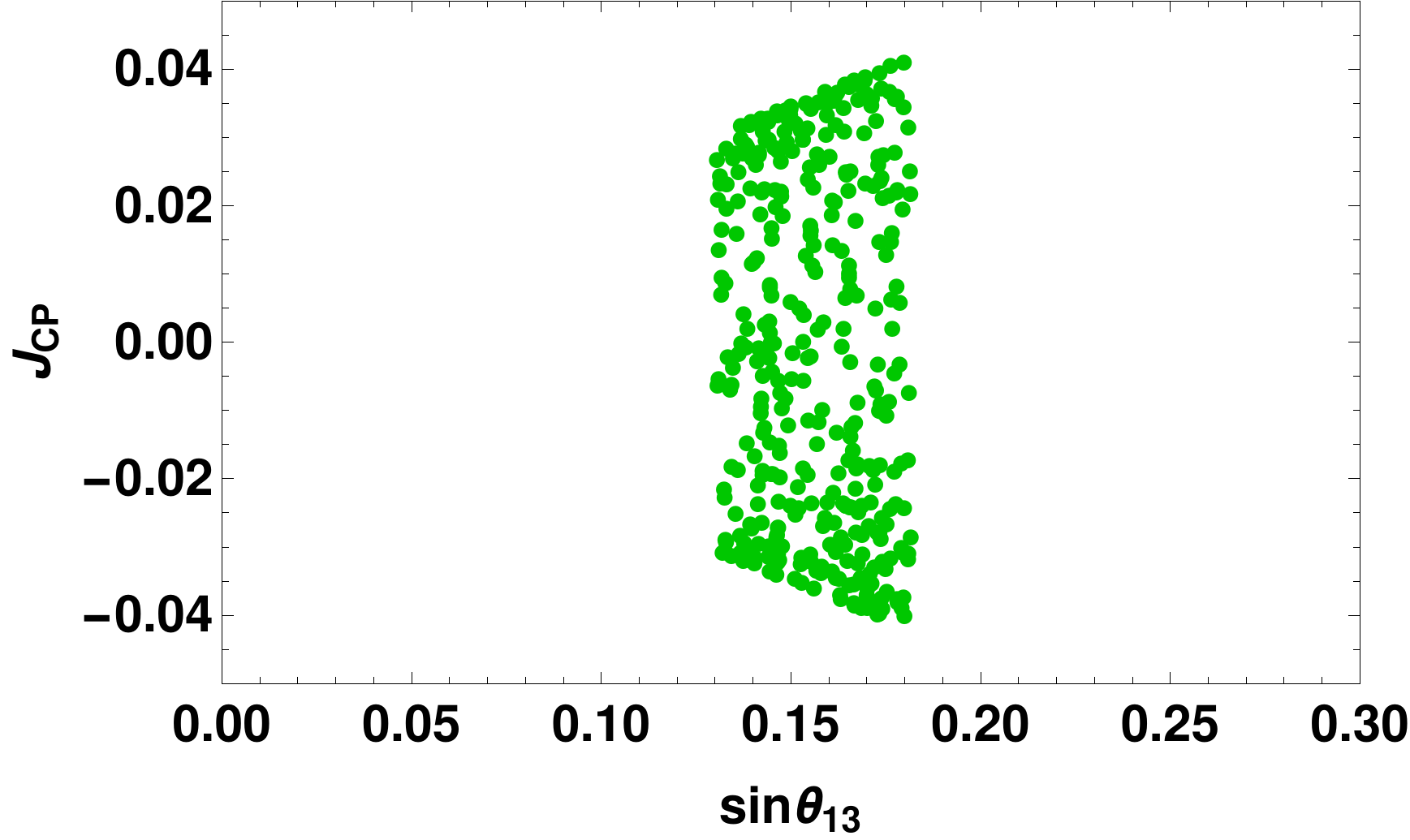}
%
	\includegraphics[width=0.48\textwidth]{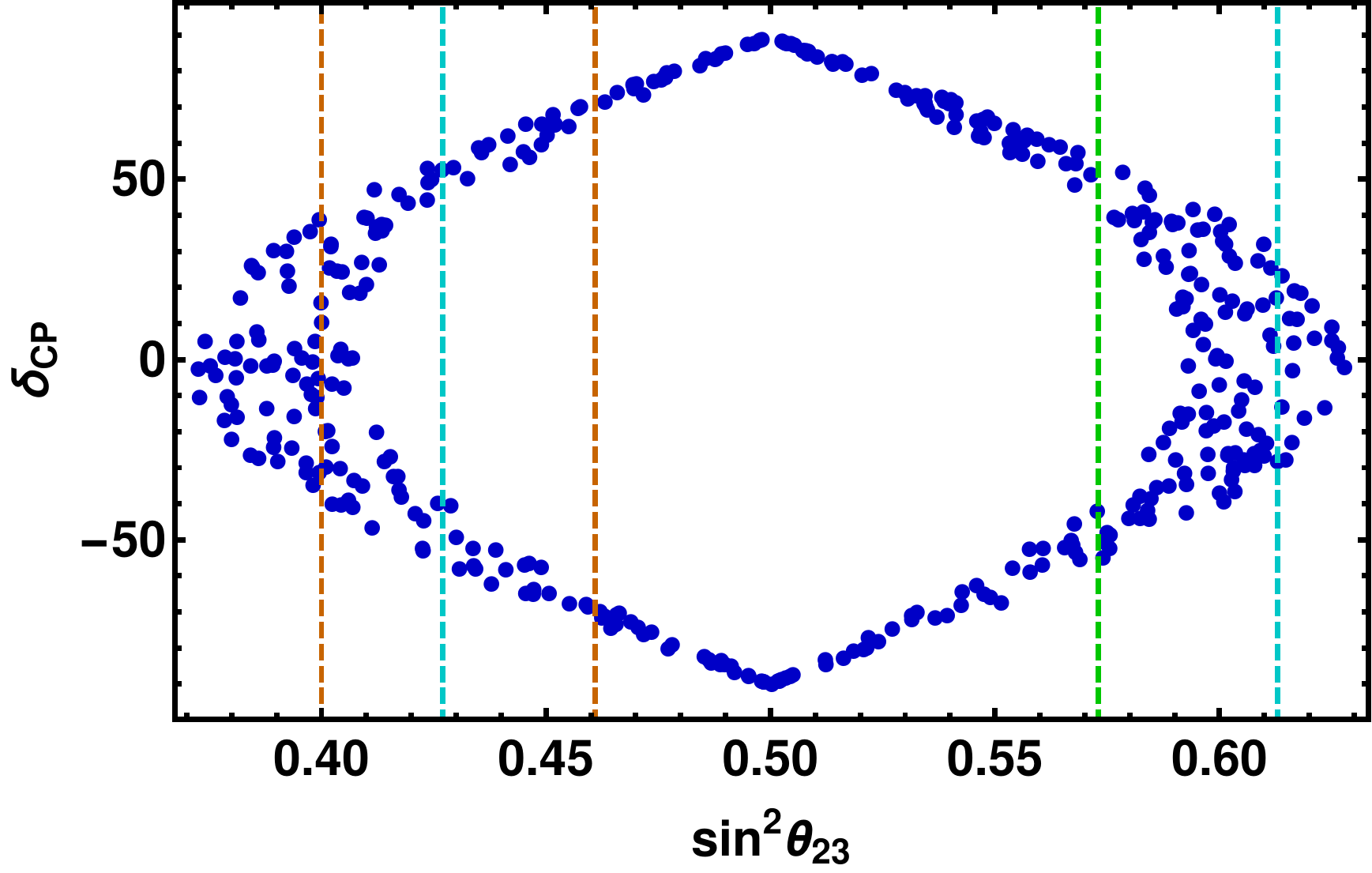}
	\includegraphics[width=0.48\textwidth]{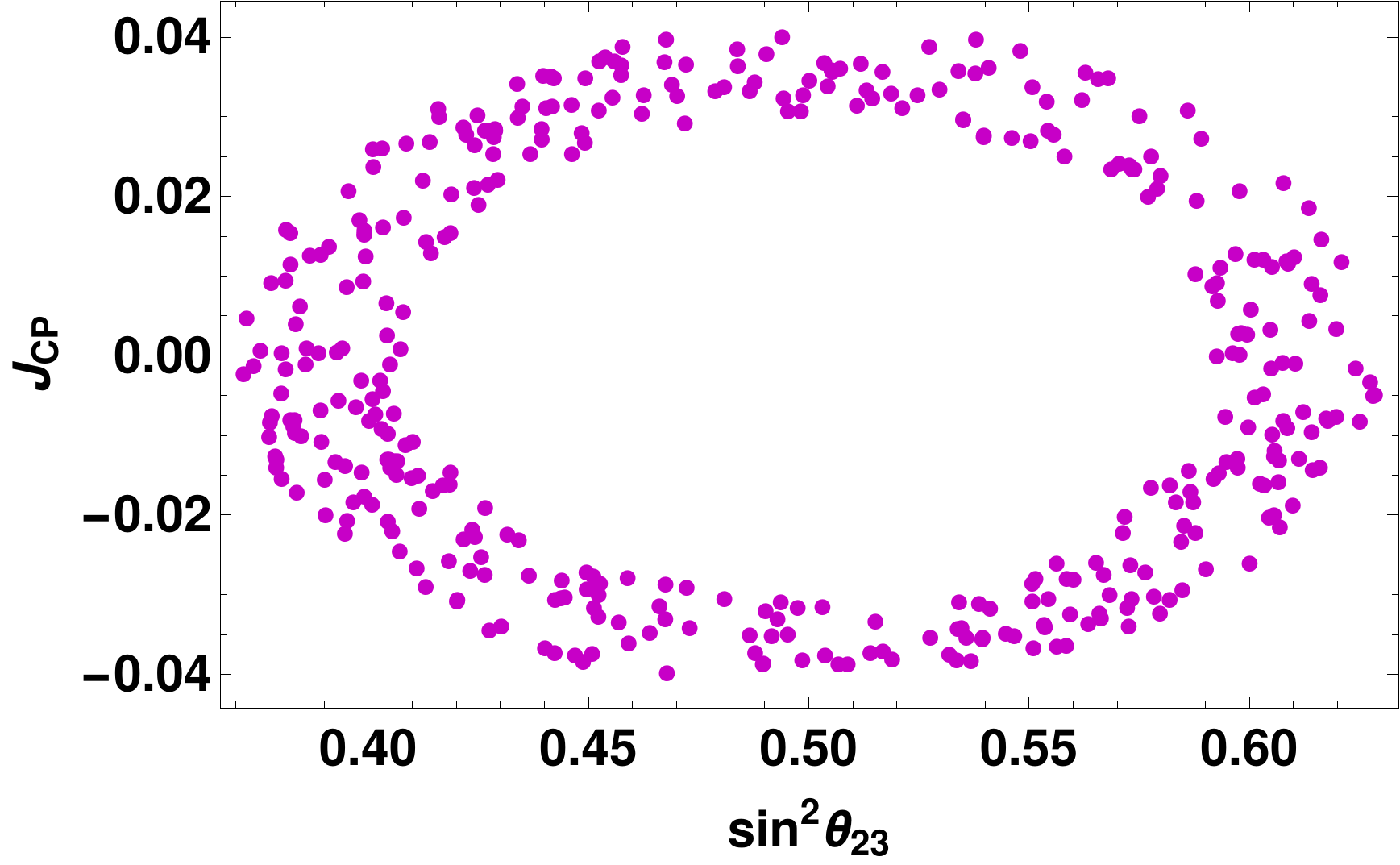}
%
	\includegraphics[width=0.48\textwidth]{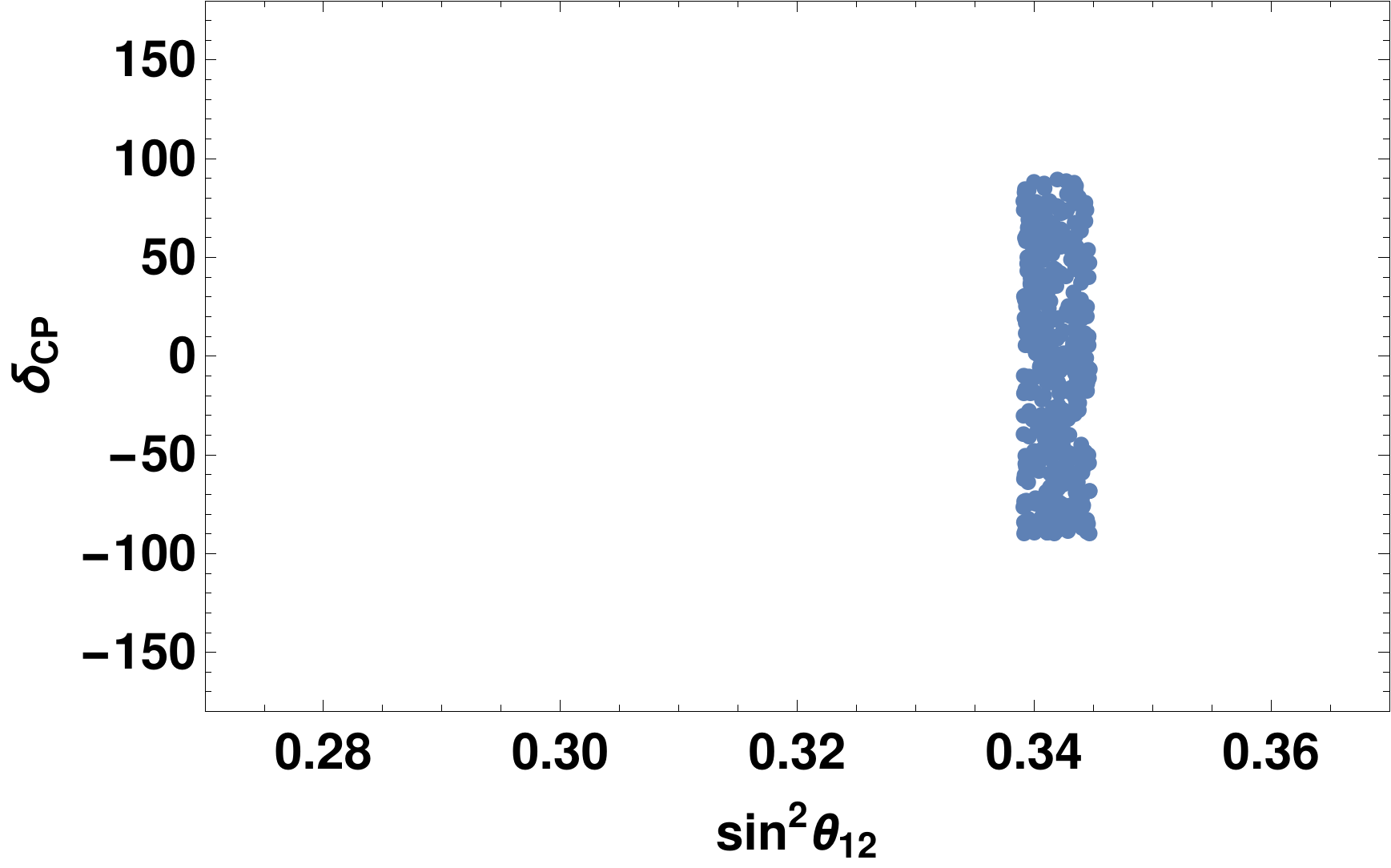}
	\includegraphics[width=0.48\textwidth]{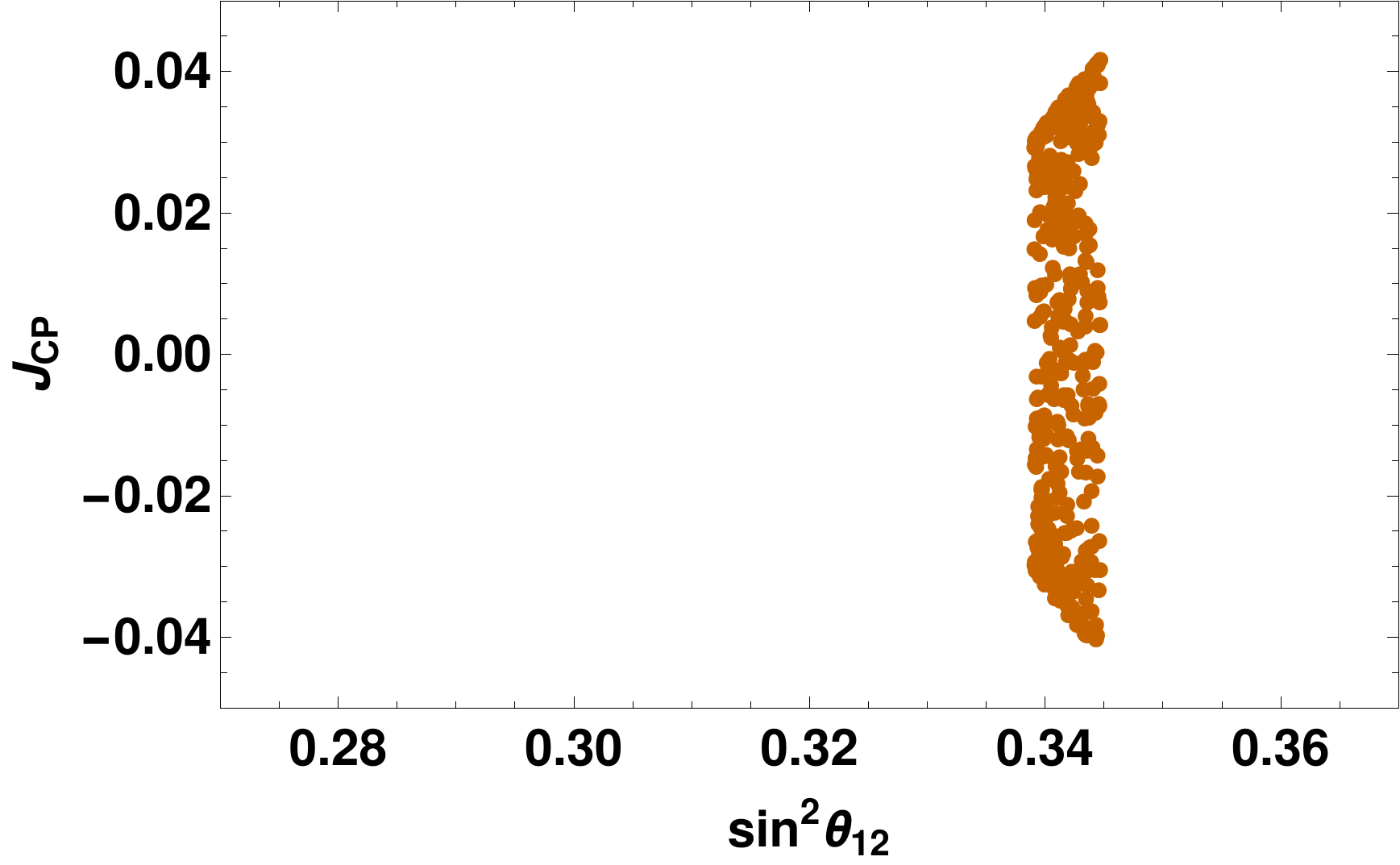}
	\caption{The plots show the inter-relation between the Dirac CP violating phase $\delta_{CP}$ and rephrasing invariant $J_{\rm CP}$  with other mixing angles $\theta_{12}$, $\theta_{23}$, $\theta_{13}$.}
	\label{fig:mixing1}
\end{figure}

Fig.\ref{fig:mixingangles-theta}, shows the variation of neutrino parameters like solar mixing angle ($\theta_{12}$), reactor mixing angle ($\theta_{13}$), 
and atmospheric mixing angle ($\theta_{23}$) along with the Jarlskog rephrasing invariant $J_{\rm CP}$ with the change of internal mixing angle $\theta$ 
. The solid green line represents $\theta_{12}$, solid blue line is for $\theta_{23}$ and 
solid magenta line is for $\theta_{13}$ while dashed lines show different predictions for rephrasing invariant $J_{\rm CP}$ by 
fixing phase $\delta=0,60,100$ for yellow,magenta and green respectively. The experimentally measured $\sigma$ allowed region for $\theta_{23}, \theta_{12}$ and $\theta_{13}$~\cite{Esteban:2016qun} 
are displayed in blue, green and magenta bands respectively. We have done random scan of internal mixing angle $\theta$ and the phase angle $\delta$ in the 
range $0-2\pi$ by fixing $M_1$ at 25~GeV, $M_2$ at 800~GeV and $M_3$ at 2~TeV,  and considered only those values which fall within the experimental range and shown the correlation between predicted neutrino oscillation 
parameters in Fig.\ref{fig:mixing1}.

\section{Non-unitarity effects in linear seesaw}
\label{sec:non unitarity}
The linear seesaw mechanism allows large mixing between light and heavy neutrinos, which gives dominant contributions to lepton flavor violating decays and Jarlskog invariants ${\bf J_{\rm CP}}$. These processes are related to non-unitarity effects in neutrino masses and mixing.\cite{Antusch:2006vwa,Antusch:2008tz,Antusch:2009pm,Forero:2011pc,FernandezMartinez:2007ms,Kanaya:1980cw,Kersten:2007vk,Malinsky:2009gw,Altarelli:2008yr,delAguila:2008hw,delAguila:2009bb,Cabrera:2019xkf} 
The complete neutral fermion spectrum with flavor and mass eigenstates are related in the following way \begin{eqnarray}
 \mid \Psi \rangle_f = \mathbb{V}^\dagger \mid \Psi \rangle_m\, \quad,  \mid \Psi \rangle_f =  \begin{pmatrix}
               \nu_L \\ \nu^c_R  \\ S_L
              \end{pmatrix} 
\quad, \mid \Psi \rangle_m  = 
              \begin{pmatrix}
               \nu_i \\ N_k
              \end{pmatrix}             
\end{eqnarray}
Here we assume $\nu_L$ with $L=e, \mu, \tau$ for flavour eigenstates, $\nu_i$ with $i=1,2,3$ for mass eigenstates, 
$\nu^c_R$ and $S_L$ as flavour eigenstates, $N_{k}$ with $k=1,2,...,6$ for mass eigenstates.  
After complete diagonalization process the physical neutral fermions are comprised of three Majorana neutrinos and three Dirac neutrinos which come up after 
six heavy neutrinos pair up. The mass formula for light as well as heavy neutrinos are given by
\begin{eqnarray}\label{mnu}
&&m_{\nu}\simeq m_{LR} m^{-1}_{RS} m^{T}_{LS} + \mbox{transpose}\nonumber \\
&&M \simeq m_{RS} + \cdots  
\end{eqnarray}

The complete $9 \times 9$ mixing matrix is of the following form\cite{Agostinho:2017wfs},
\begin{eqnarray}
 \mathbb{V} = \mathbb{W} \cdot \mathbb{U} 
    = \begin{pmatrix} \sqrt{1+X X^\dagger} U_\nu & X\,U_N \nonumber \\ X^\dagger\,U_\nu & \sqrt{1+X^\dagger X} U_N 
      \end{pmatrix} \cdot \begin{pmatrix} U_\nu & 0 \\ 0 & U_N
\end{pmatrix}
\end{eqnarray}

\begin{figure}[t] 
  \begin{minipage}[b]{0.5\linewidth}
    \includegraphics[width=.80\linewidth]{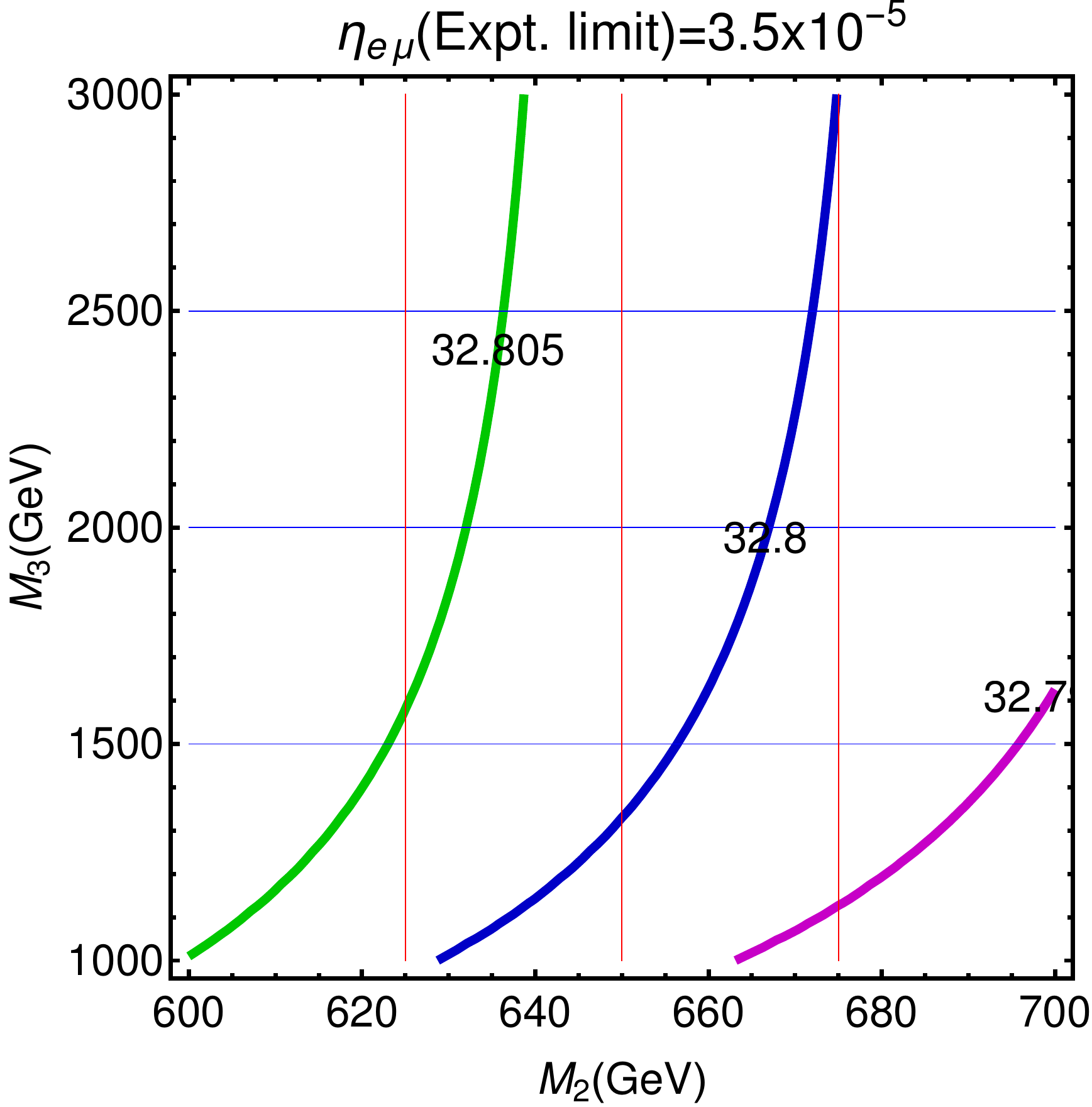} 
  \end{minipage} 
  \begin{minipage}[b]{0.5\linewidth}
    \includegraphics[width=.80\linewidth]{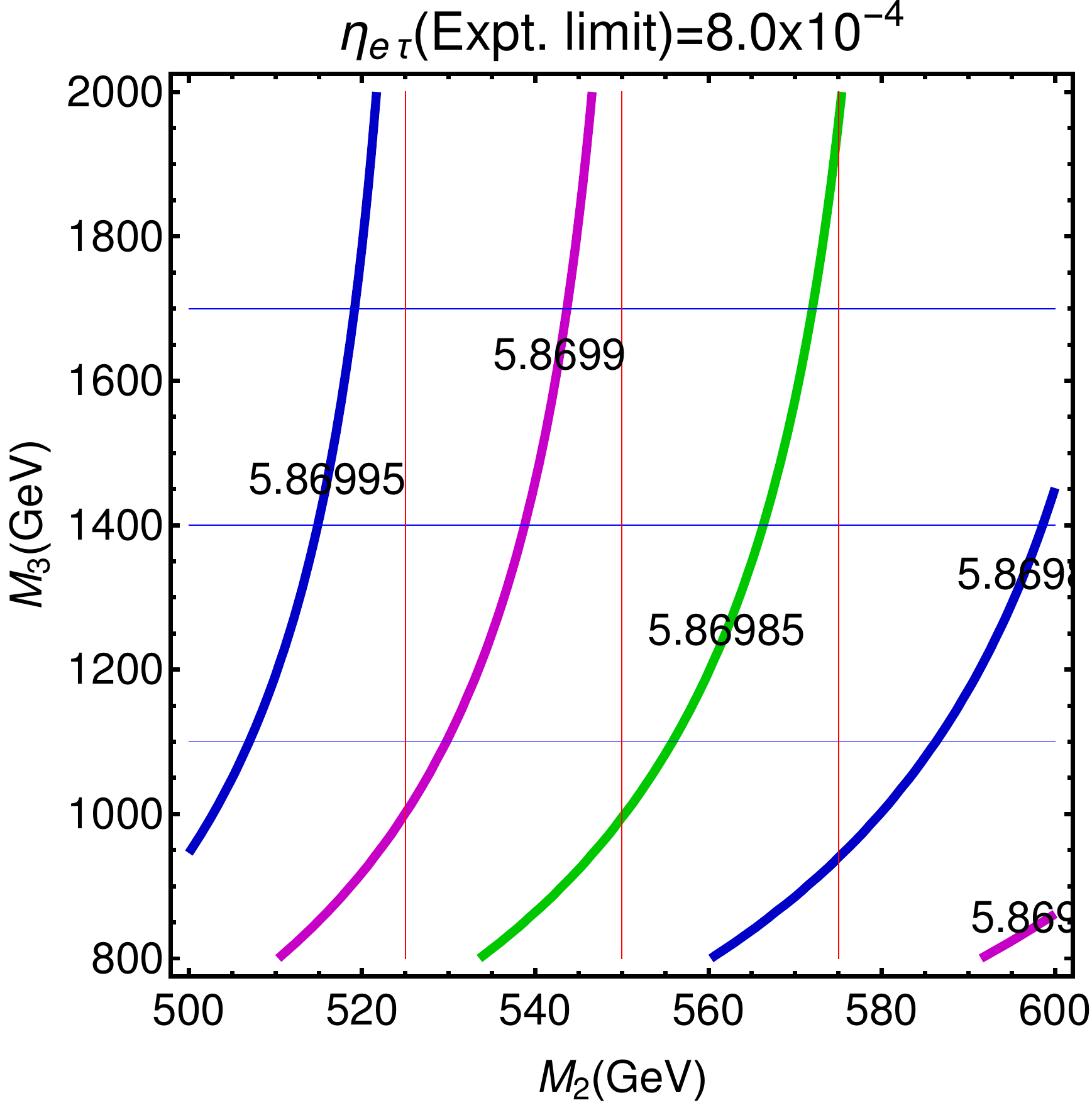} 
  \end{minipage} 
   \begin{minipage}[b]{0.5\linewidth}
    \includegraphics[width=.80\linewidth]{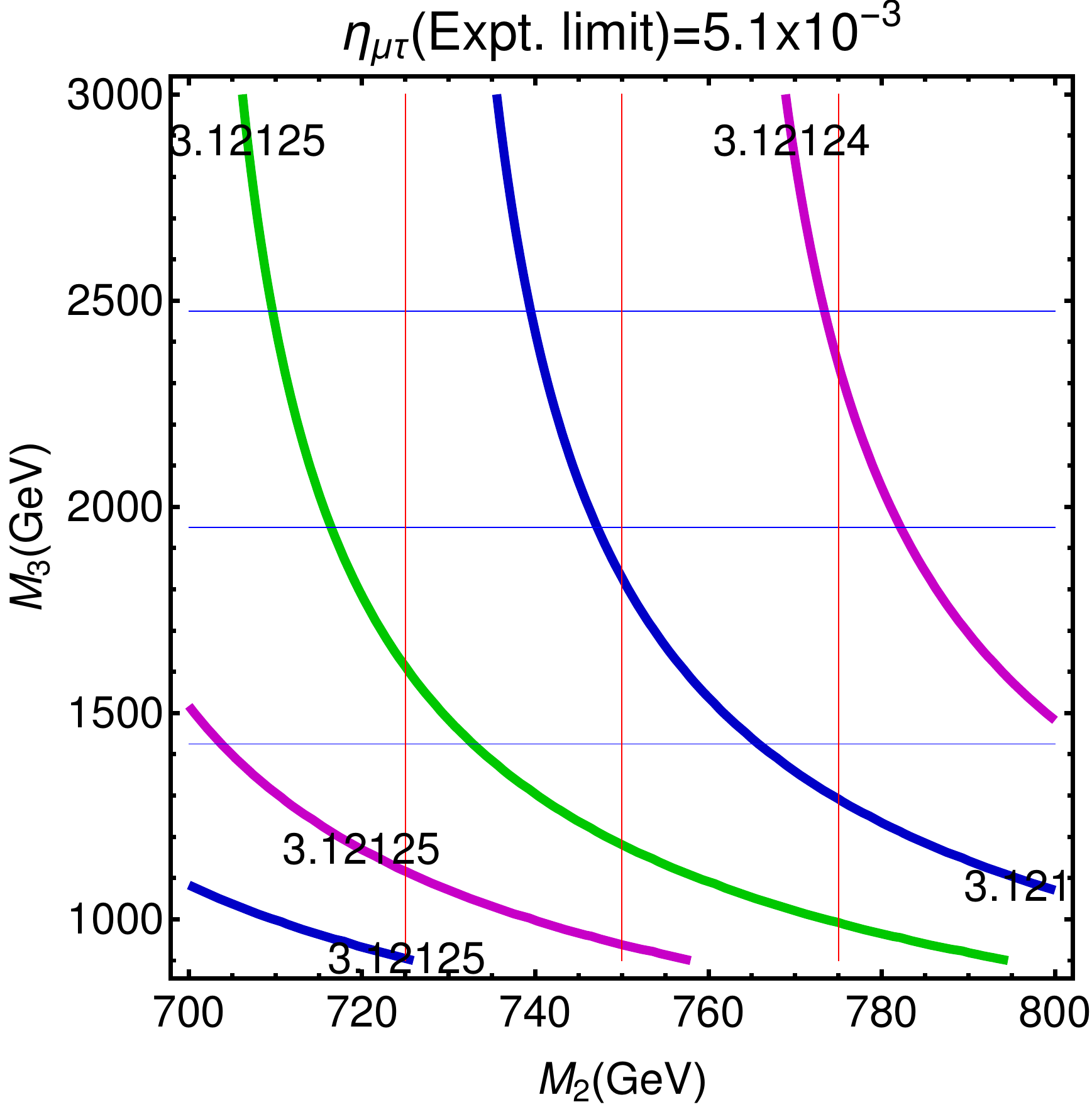} 
  \end{minipage}
  \begin{minipage}[b]{0.5\linewidth}
    \includegraphics[width=.80\linewidth]{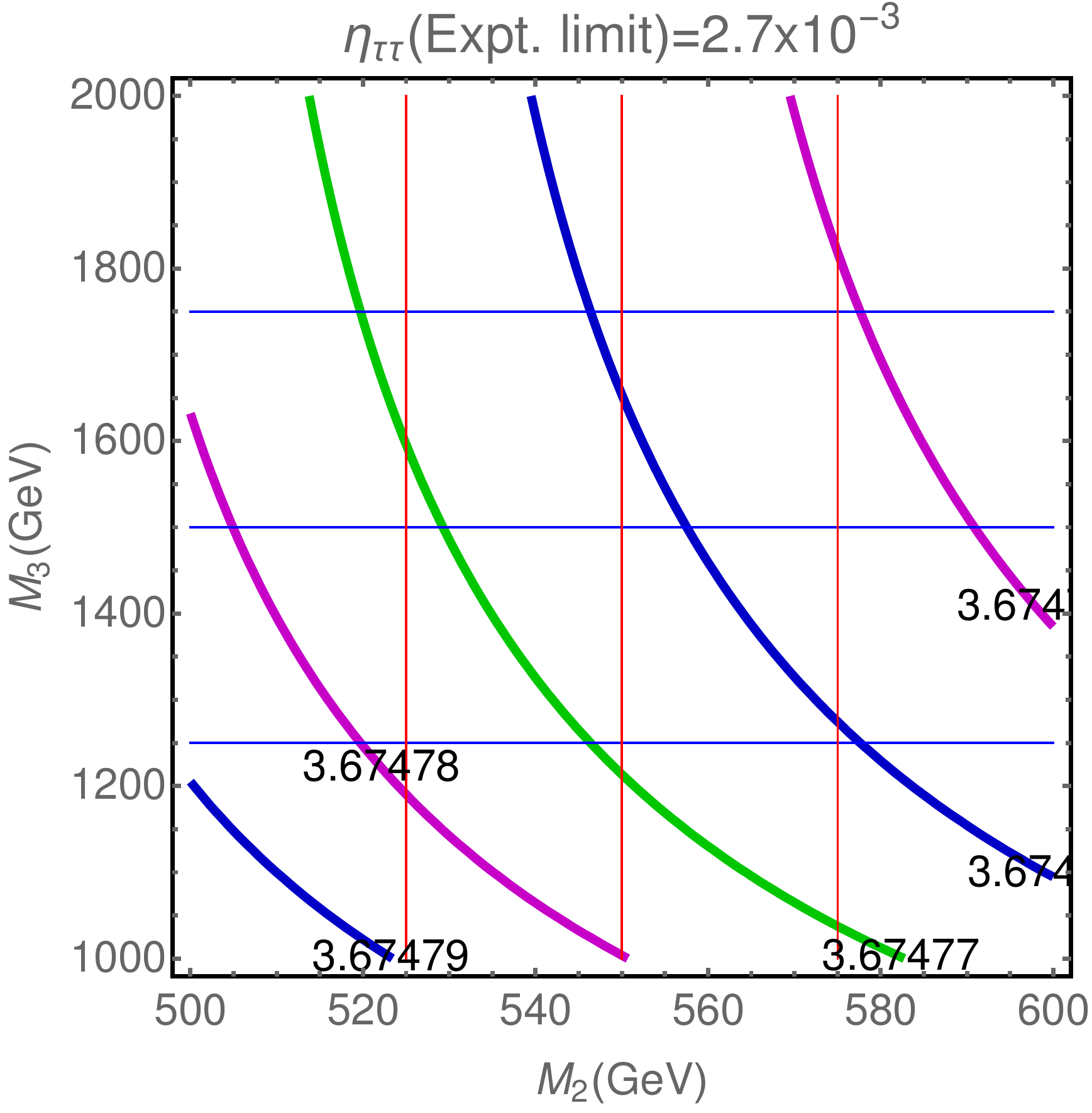} 
  \end{minipage} 
\caption{Contours plots in the plane of $M_2$ and $M_3$ for different fixed values of $M_1$ by saturating the experimental values of 
unitarity violating parameter $\eta$ in $e\mu, e\tau, \mu\tau, \tau \tau$ sectors.} 
\label{fig:eta-contour}
\end{figure}

The unitary mixing matrices $U_\nu$ and $U_N$ are required to diagonalise the light neutrino mass matrix $m_\nu$ and heavy neutrino 
mass matrix $M$. In usual case, the light active Majorana neutrino mass matrix is diagonalized by the PMNS mixing 
matrix $U_{\rm PMNS}$ as $U_{\rm PMNS}^{\dagger}\, m_{\nu}\, U^*_{\rm PMNS} = \text{diag}
\left(m_{1},m_{2},m_{3}\right)$ where $m_1,m_2,m_3$ are mass eigenvalues for light neutrinos. 
However, due to the presence of extra heavy neutrinos, the diagonalizing mixing matrix in case of linear 
seesaw mechanism ~\cite{Hettmansperger:2011bt,FernandezMartinez:2007ms,Antusch:2009pm,Forero:2011pc}(where the neutral 
lepton sector comprises of light active Majorana neutrinos plus two right-handed sterile neutrinos) is no longer unitary 
and is given by,
\begin{equation} 
\mathbb{N} = \left(1-\eta \right) U_{\nu} \equiv  \left(1-\eta \right) U_{\rm PMNS}\, , 
\end{equation}
Here,$\eta$ is a measure of deviation from unitarity in the PMNS mixing matrix in the light neutrino sector. The non-unitarity effect can be recast in terms of  $m_{LR}$ and $m_{RS}$ as~\cite{Forero:2011pc},
\begin{equation}
\eta=\frac{1}{2}m_{LR}^*{m_{RS}^{\dagger~ -1}} m_{RS}^{-1}m_{LR}^T\;.\label{et}
\end{equation} 

In linear seesaw scheme invoked with $A_4$ flavor symmetry, the structure of $m_{LR}$ and $m_{LS}$ are proportional to the identity matrix 
and the other matrix $m_{RS}$ is diagonalized in the following way
\begin{equation}
m^d_{RS} = \Big(U_{\rm TBM} U_{13} \Big)^T m_{RS} \Big(U_{\rm TBM} U_{13} \Big) 
\end{equation} 
where,
\begin{eqnarray}
&&U_{\rm TBM} U_{13} = \begin{pmatrix}
 \frac{2}{\sqrt{6}}  \cos\theta & \frac{1}{\sqrt{3}} & \frac{2}{\sqrt{6}} \sin\theta e^{-i \delta} \nonumber \\
 -\frac{1}{\sqrt{6}} \cos\theta+\frac{1}{\sqrt{2}}  \sin\theta e^{i \delta} & \frac{1}{\sqrt{3}} & -\frac{1}{\sqrt{6}} \sin\theta e^{-i \delta}-\frac{1}{\sqrt{2}}  \cos\theta \\
 -\frac{1}{\sqrt{6}}  \cos\theta-\frac{1}{\sqrt{2}}  \sin\theta e^{i \delta} & \frac{1}{\sqrt{3}} & -\frac{1}{\sqrt{6}}  \sin\theta e^{-i\delta }+\frac{1}{\sqrt{2}}  \cos\theta 
  \end{pmatrix}
\nonumber \\
&& 
m^d_{RS} = \begin{pmatrix}
M_1 & 0 & 0   \\
0      & M_2  & 0 \\
0 & 0 & M_3
\end{pmatrix}
\end{eqnarray}

 
 \begin{figure}[t] 
  \begin{minipage}[b]{0.5\linewidth}
    \includegraphics[width=.80\linewidth]{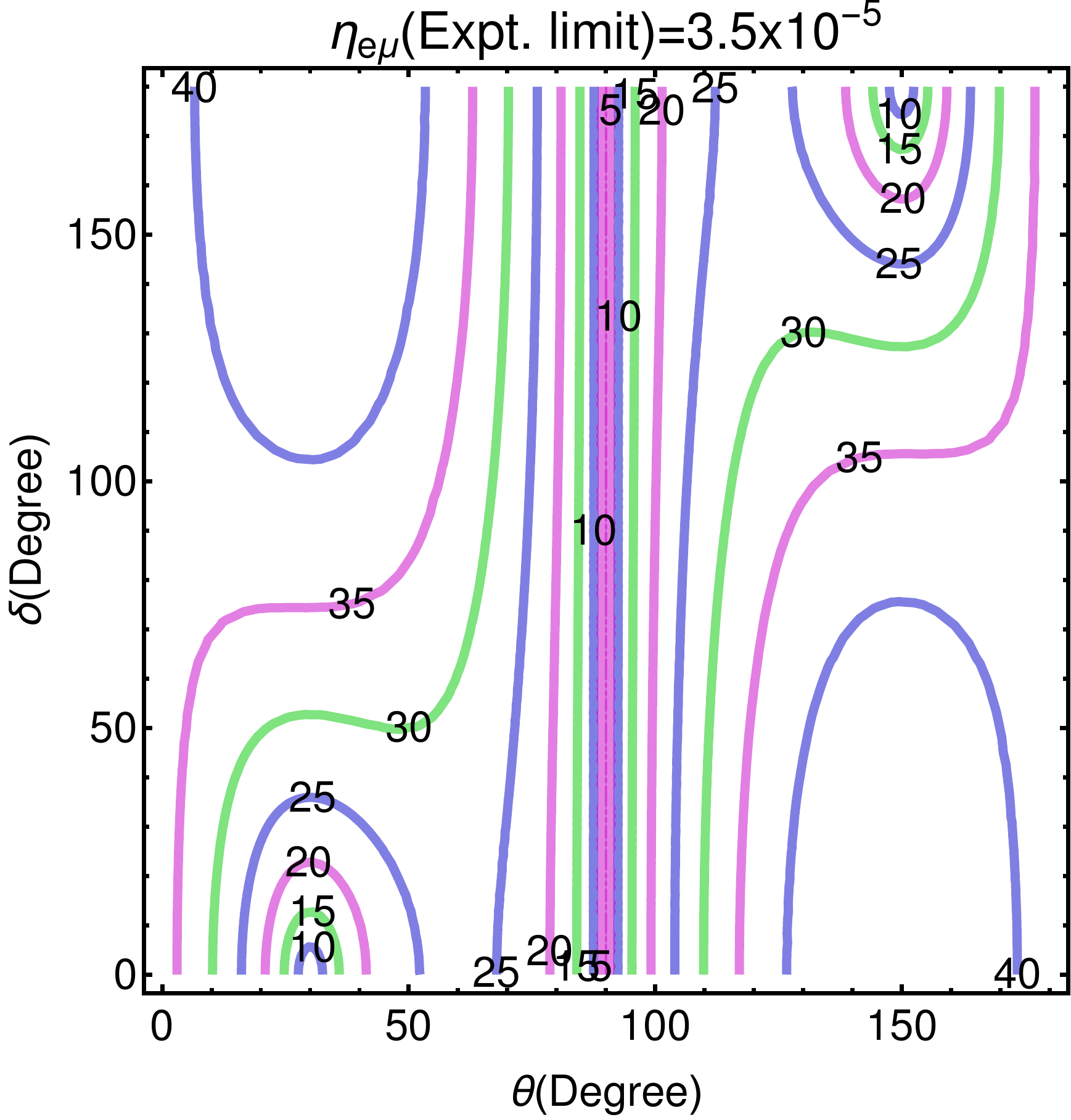} 
  \end{minipage} 
  \begin{minipage}[b]{0.5\linewidth}
    \includegraphics[width=.80\linewidth]{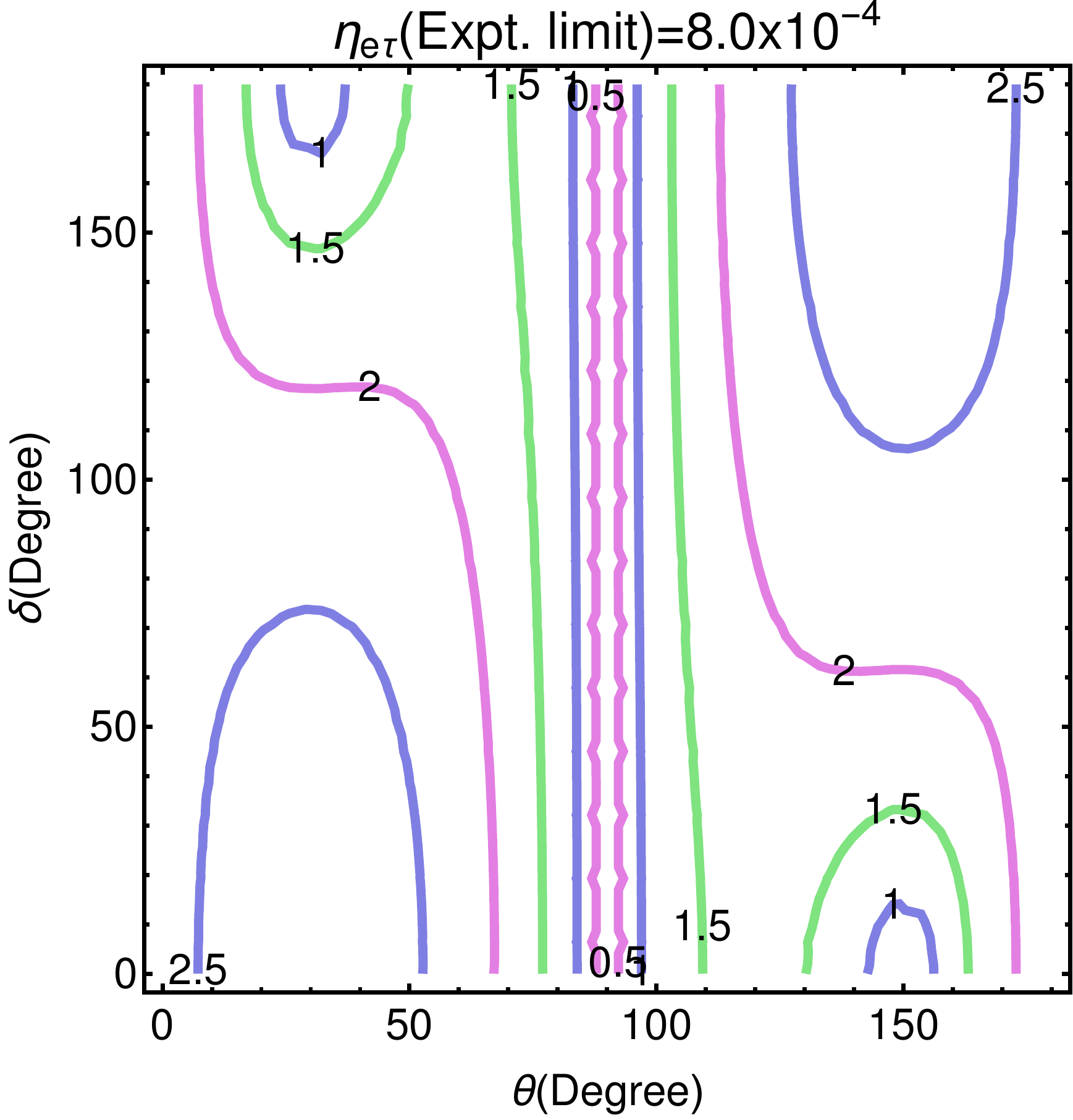} 
  \end{minipage} 
   \begin{minipage}[b]{0.5\linewidth}
    \includegraphics[width=.80\linewidth]{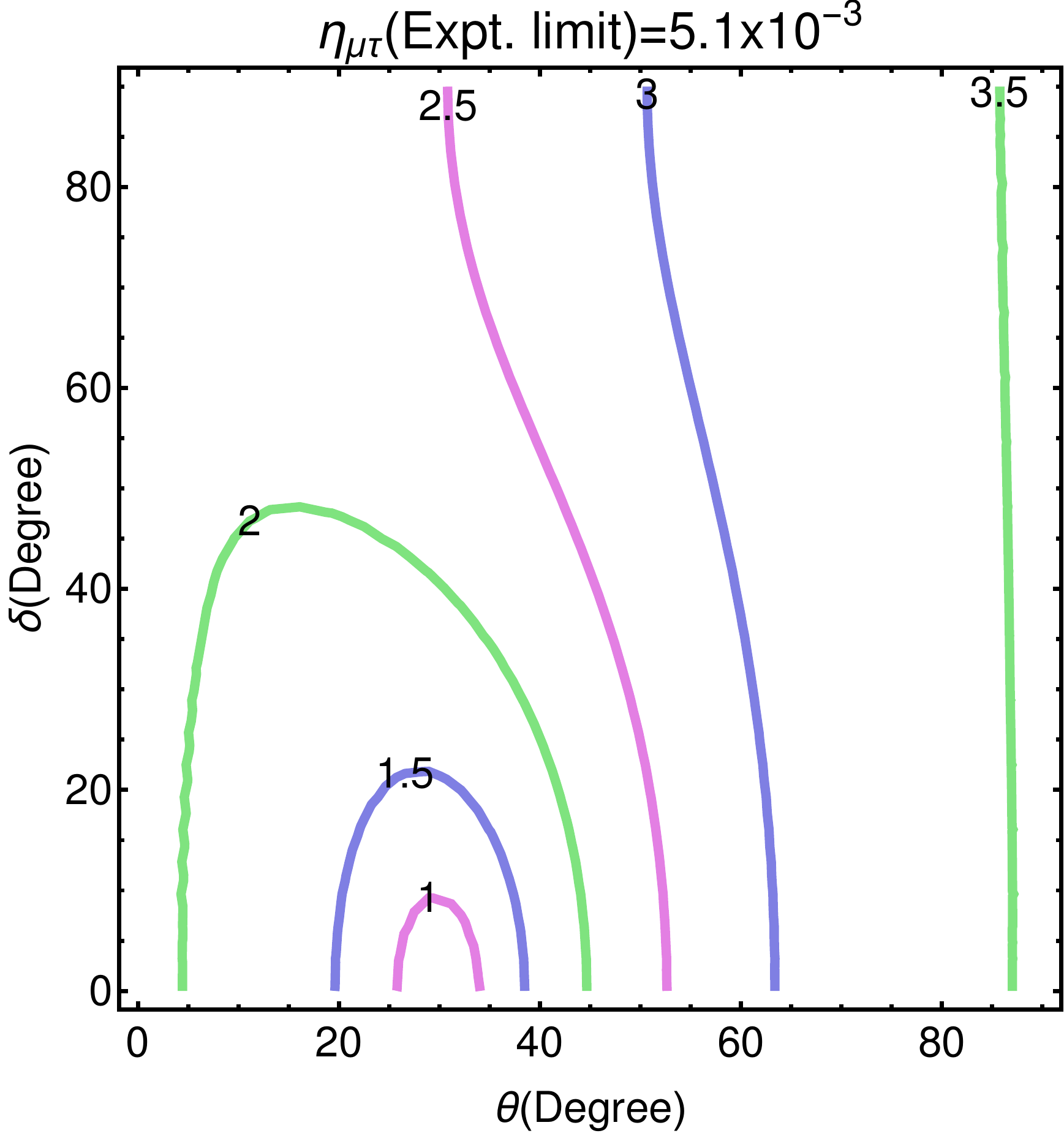} 
  \end{minipage}
  \begin{minipage}[b]{0.5\linewidth}
    \includegraphics[width=.80\linewidth]{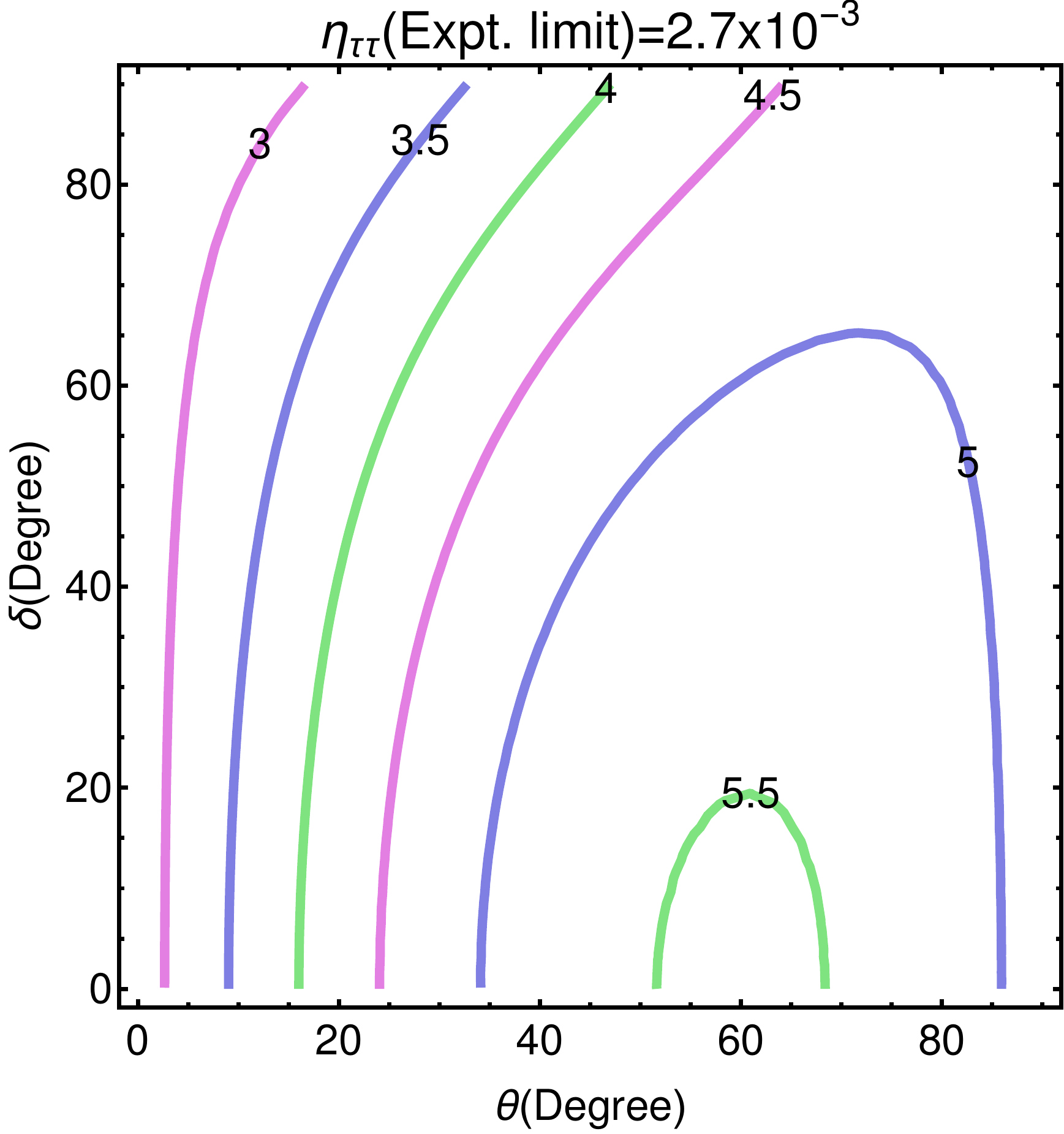} 
  \end{minipage} 
\caption{Correlation plot between the internal mixing angle $\theta$ and phase $\delta$ for observable unitarity effects 
at current and planned experiments  in $e\mu, e\tau, \mu\tau, \tau \tau$ sectors.} 
\label{fig:delta-contour}
\end{figure}
 
 The linear seesaw formula provides sub-eV scale (of order 0.1~eV) masses for light active neutrinos with values of 
 model parameters, $m_{LR} \sim 0.5$~GeV, $m_{RS} \sim 10^3$ GeV and $m_{LS} \sim 100$~eV. 
 The unitarity violation in $e\mu$,$e\tau$,$\mu \tau$, $\tau\tau$ sector can be expressed by saturating the experimental 
bound which are $|\eta_{e \mu}| < 3.5\times 10^{-5}$, $|\eta_{e\tau}| < 8.0\times 10^{-4}$, $|\eta_{\mu \tau}| < 5.1\times 10^{-3}$ 
and $|\eta_{\tau\tau}| < 2.7\times 10^{-3}$ \cite{Awasthi:2013ff,LalAwasthi:2011aa,Miranda:2019ynh} as
\begin{eqnarray}
\pmb{\eta_{e\mu}}&=& m^2_d \times \Bigg[ 
\frac{1}{M^2_1} \Bigg(\sqrt{\frac{2}{3}} \cos \theta \bigg( 
\frac{-\cos \theta}{\sqrt{6}} + \frac{\cos\delta \sin\theta}{\sqrt{2}}
+ \frac{i \sin\delta \sin\theta}{\sqrt{2}}
\bigg)
  \Bigg)
    + \frac{1}{3 M^2_2} \nonumber \\
&&\hspace*{-0.5cm}+\frac{1}{M^2_3} \Big(\sqrt{\frac{2}{3}} \cos \delta \sin\theta -  
i \sqrt{\frac{2}{3}} \sin\delta \sin\theta \Big) \Big(
\frac{-\cos \theta}{\sqrt{2}} - \frac{\cos\delta \sin\theta}{\sqrt{6}}
+ \frac{i \sin\delta \sin\theta}{\sqrt{6}}
\Big)
\Bigg]
\label{eq:eta_emu}
\end{eqnarray}
\noindent
\begin{eqnarray}
\pmb{\eta_{e\tau}}&=& m^2_d \times \Bigg[ 
\frac{1}{M^2_1} \Bigg(\sqrt{\frac{2}{3}} \cos \theta \bigg( 
\frac{-\cos \theta}{\sqrt{6}} - \frac{\cos\delta \sin\theta}{\sqrt{2}}
+ \frac{i \sin\delta \sin\theta}{\sqrt{2}}
\bigg)
  \Bigg)
    + \frac{1}{3 M^2_2} \nonumber \\
&&\hspace*{-0.5cm}+\frac{1}{M^2_3} \Big(\sqrt{\frac{2}{3}} \cos \delta \sin\theta -  
i \sqrt{\frac{2}{3}} \sin\delta \sin\theta \Big) \Big(
\frac{\cos \theta}{\sqrt{2}} - \frac{\cos\delta \sin\theta}{\sqrt{6}}
- \frac{i \sin\delta \sin\theta}{\sqrt{6}}
\Big)
\Bigg]
\label{eq:eta_etau}
\end{eqnarray}
\begin{eqnarray}
\pmb{\eta_{\mu\tau}}&=& m^2_d \times \Bigg[ 
\frac{1}{M^2_1} \Bigg(\bigg( 
\frac{-\cos \theta}{\sqrt{6}} - \frac{\cos\delta \sin\theta}{\sqrt{2}}
+ \frac{i \sin\delta \sin\theta}{\sqrt{2}}
\bigg)\bigg( 
\frac{-\cos \theta}{\sqrt{6}} + \frac{\cos\delta \sin\theta}{\sqrt{2}}
+ \frac{i \sin\delta \sin\theta}{\sqrt{2}}
\bigg)
  \Bigg)
    + \frac{1}{3 M^2_2} \nonumber \\
&&\hspace*{-0.5cm}+\frac{1}{M^2_3}  \Big(
\frac{\cos \theta}{\sqrt{2}} - \frac{\cos\delta \sin\theta}{\sqrt{6}}
- \frac{i \sin\delta \sin\theta}{\sqrt{6}}
\Big)
\Big(
\frac{-\cos \theta}{\sqrt{2}} - \frac{\cos\delta \sin\theta}{\sqrt{6}}
+\frac{i \sin\delta \sin\theta}{\sqrt{6}}\Big)
\Bigg]
\label{eq:eta_mutau}
\end{eqnarray}
\begin{eqnarray}
\pmb{\eta_{\tau\tau}}&=& m^2_d \times \Bigg[ 
\frac{1}{M^2_1} \Bigg(\bigg( 
\frac{-\cos \theta}{\sqrt{6}} - \frac{\cos\delta \sin\theta}{\sqrt{2}}
- \frac{i \sin\delta \sin\theta}{\sqrt{2}}
\bigg)\bigg( 
\frac{-\cos \theta}{\sqrt{6}} - \frac{\cos\delta \sin\theta}{\sqrt{2}}
+ \frac{i \sin\delta \sin\theta}{\sqrt{2}}
\bigg)
  \Bigg)
    + \frac{1}{3 M^2_2} \nonumber \\
&&\hspace*{-0.5cm}+\frac{1}{M^2_3}  \Big(
\frac{\cos \theta}{\sqrt{2}} - \frac{\cos\delta \sin\theta}{\sqrt{6}}
- \frac{i \sin\delta \sin\theta}{\sqrt{6}}
\Big)
\Big(
\frac{\cos \theta}{\sqrt{2}} - \frac{\cos\delta \sin\theta}{\sqrt{6}}
+\frac{i \sin\delta \sin\theta}{\sqrt{6}}\Big)
\Bigg]
\label{eq:eta_tautau}
\end{eqnarray}

In Fig.\ref{fig:eta-contour} we have used experimental values of $\eta$ in the 
$e\mu$,$e\tau$,$\mu\tau$,$e\tau\tau$ sectors \cite{Awasthi:2013ff,LalAwasthi:2011aa,Miranda:2019ynh} and plotted $M_3$ 
versus $M_2$  where the curves show the allowed values of $M_1$. We have used the equations \ref{eq:eta_emu}, \ref{eq:eta_etau}, \ref{eq:eta_mutau}, 
\ref{eq:eta_tautau} for the four plots and set $\theta=120$ degree, $\delta=60$ degree. Whereas in Fig.\ref{fig:delta-contour} we have 
fixed $M_2$ at 100 GeV, $M_3$ at 2 TeV and plotted $\delta$ versus $\theta$ for observable $\eta$ at experiments.
 
\section{Low Energy Lepton Flavour Violating Processes}
\label{sec:LFV}

The observation of neutrino oscillations strongly hints that lepton flavor violation might be occuring in other processes as well. 
In our model, the mechanism of Majorana neutrino mass generation is associated with the occurrence of 
charged lepton flavor violation (LFV). LFV is highly suppressed by GIM mechanism, that is, $(\Delta m^2_\nu/m_W^2) \approx 10^{-50}$ and is 
well below any experimental sensitivity in case only light neutrinos contribute to them. However, in the considered left-right symmetric framework due to the 
contribution from heavy right-handed neutrinos sizable charged lepton flavor violation occurs. For a discussion we focus here on low energy LFV processes 
$\mu\to e\gamma$, $\mu\to eee$ and $\mu\to e$~conversion in nuclei because of their sensitivity and omit LFV $\tau$ decays.For a review of LFV and new physics scenarios, 
one may refer \cite{Deppisch:2012vj}. 

 \begin{figure}[t] 
 \centering
  \begin{minipage}[b]{0.89\linewidth}
    \includegraphics[width=.80\linewidth]{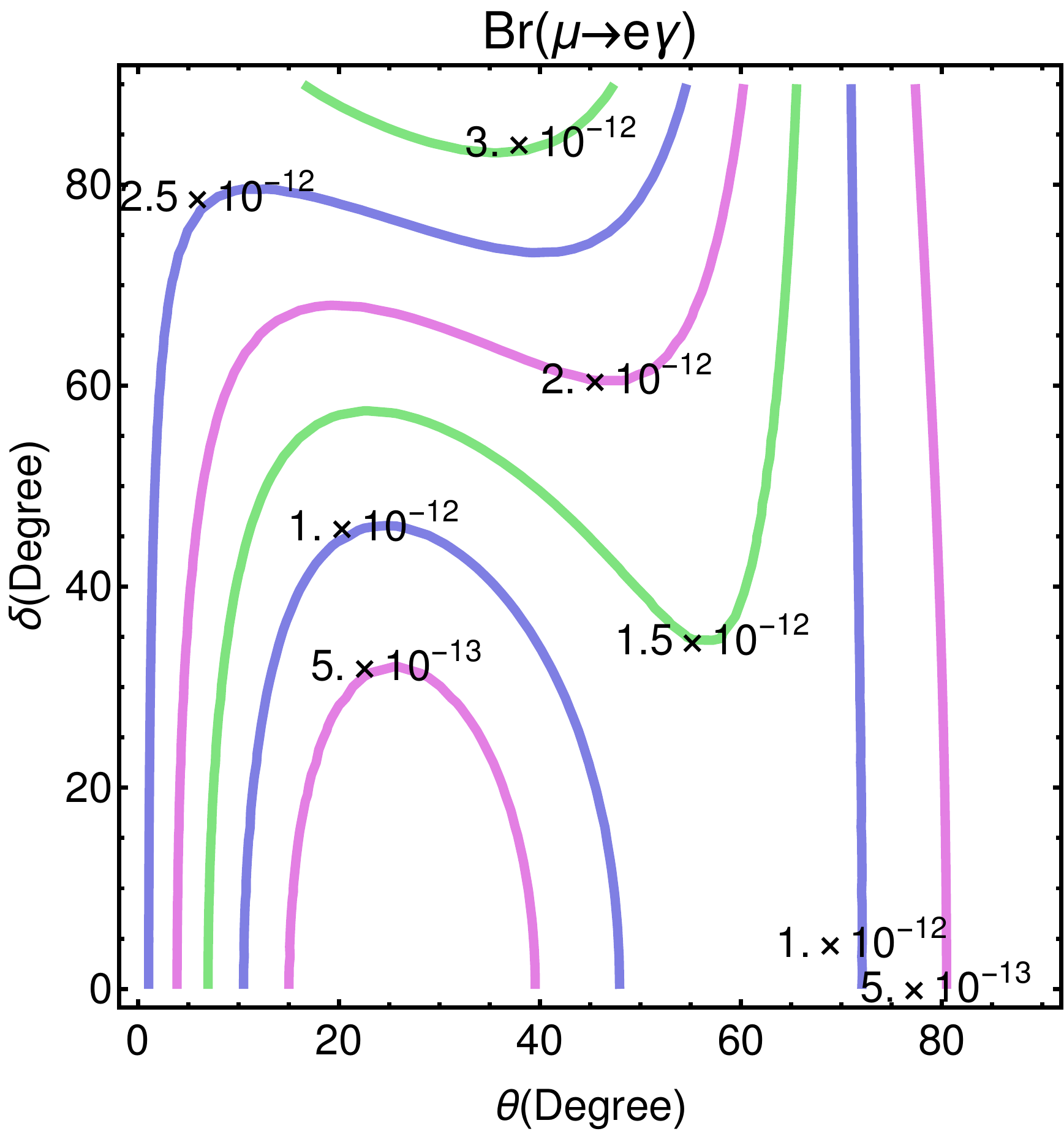} 
  \end{minipage}
\caption{Correlation plot between the internal mixing angle $\theta$ and phase $\delta$ for branching ratios for the 
lepton flavour violating processes, ${\rm BR}(\mu \to e \gamma)$.}
\label{fig:branchingR1}
\end{figure}
As discussed in the previous section, unitarity violation has implications on prediction for lepton flavor violation. Since the measure of unitarity violation is of the order 
of $M_{LR}^2/M^2$, $\mu \to e\gamma$ term plays a vital role  in deriving constraints on input parameters like internal mixing angle $\theta$ and phases $\delta$. 
The branching ratio for this particular process $\mu \to e \gamma$ is given by \cite{Ibarra:2011xn}
\begin{align}
	{\rm BR}(\mu \to e \gamma) = 
		\frac{3\alpha}{32\pi} \sum_{i=1}^3 f\left(\frac{M_i}{M_W}\right) 
		\left| \Theta_{\mu i}^\ast \, \Theta _{e i}\right|^2 \,,
\end{align}
Here, $M_i$ denotes for physical masses for pseudo-Dirac neutrinos, the other loop factor $f(M^2_i/M^2_W)$ is the order of one and this results,
\begin{align}
	{\rm BR}(\mu \to e \gamma) \simeq 
		8.4 \times 10^{-14} \cdot
		\left(\frac{|(\Theta\Theta^\dagger)_{e\mu}|}{10^{-5}}\right)^2\,.
\end{align}
We examined how the input model parameters are correlated by saturating the experimental bounds on these LFV processes.
The term $\Theta_{\alpha i}\Theta_{\beta i}^\dagger \simeq \pmb{\eta_{\alpha\beta}}$ in the above equation represents deviation of unitarity in the 
lepton sector which has been discussed in previous section . It has also been demonstrated in contour plots in the plane of internal mixing angle $\theta$ and phase $\delta$ in Fig.\ref{fig:branchingR1}.

 \begin{figure}[t] 
  \begin{minipage}[b]{0.5\linewidth}
    \includegraphics[width=.80\linewidth]{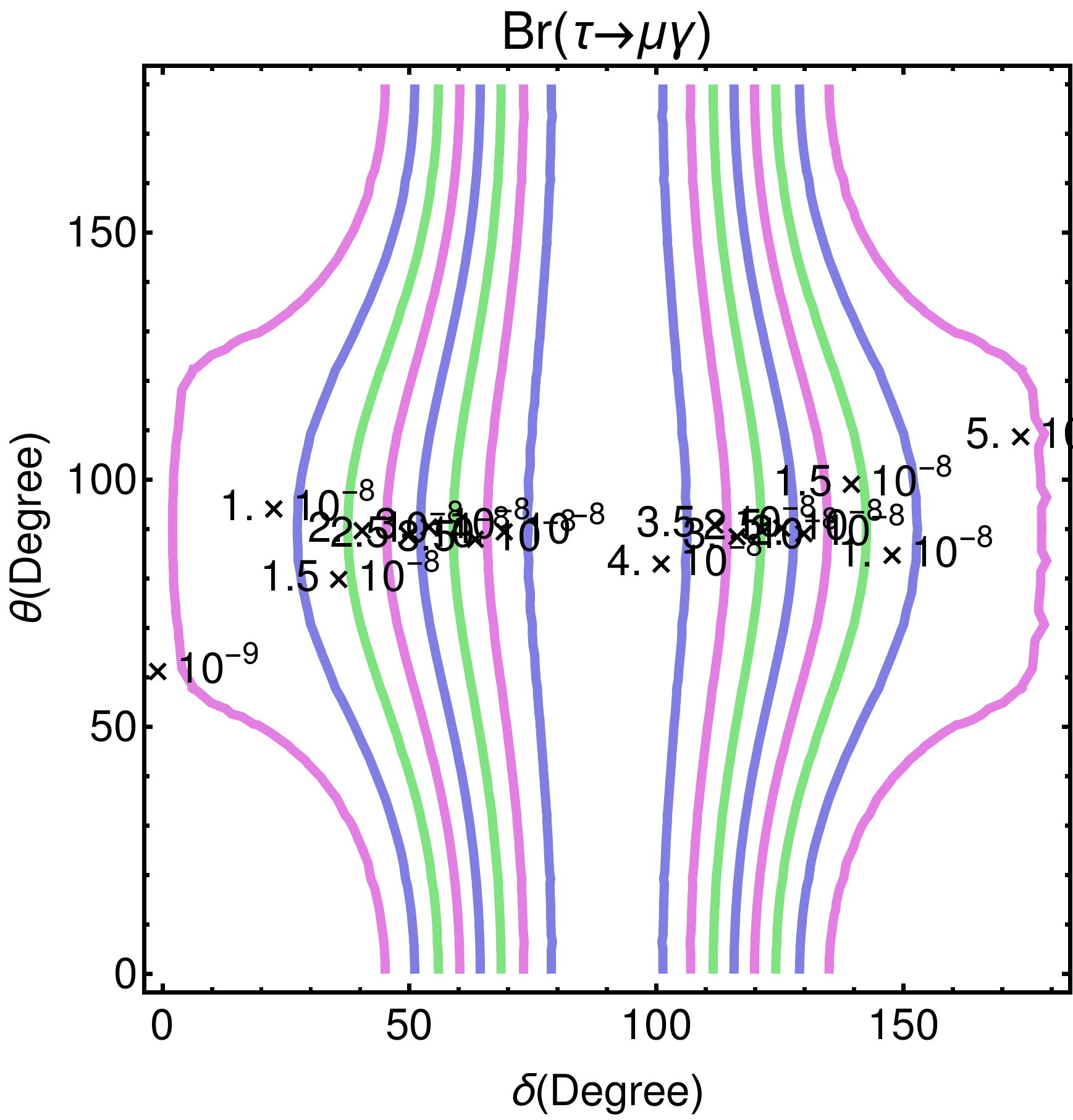} 
  \end{minipage} 
  \begin{minipage}[b]{0.5\linewidth}
    \includegraphics[width=.80\linewidth]{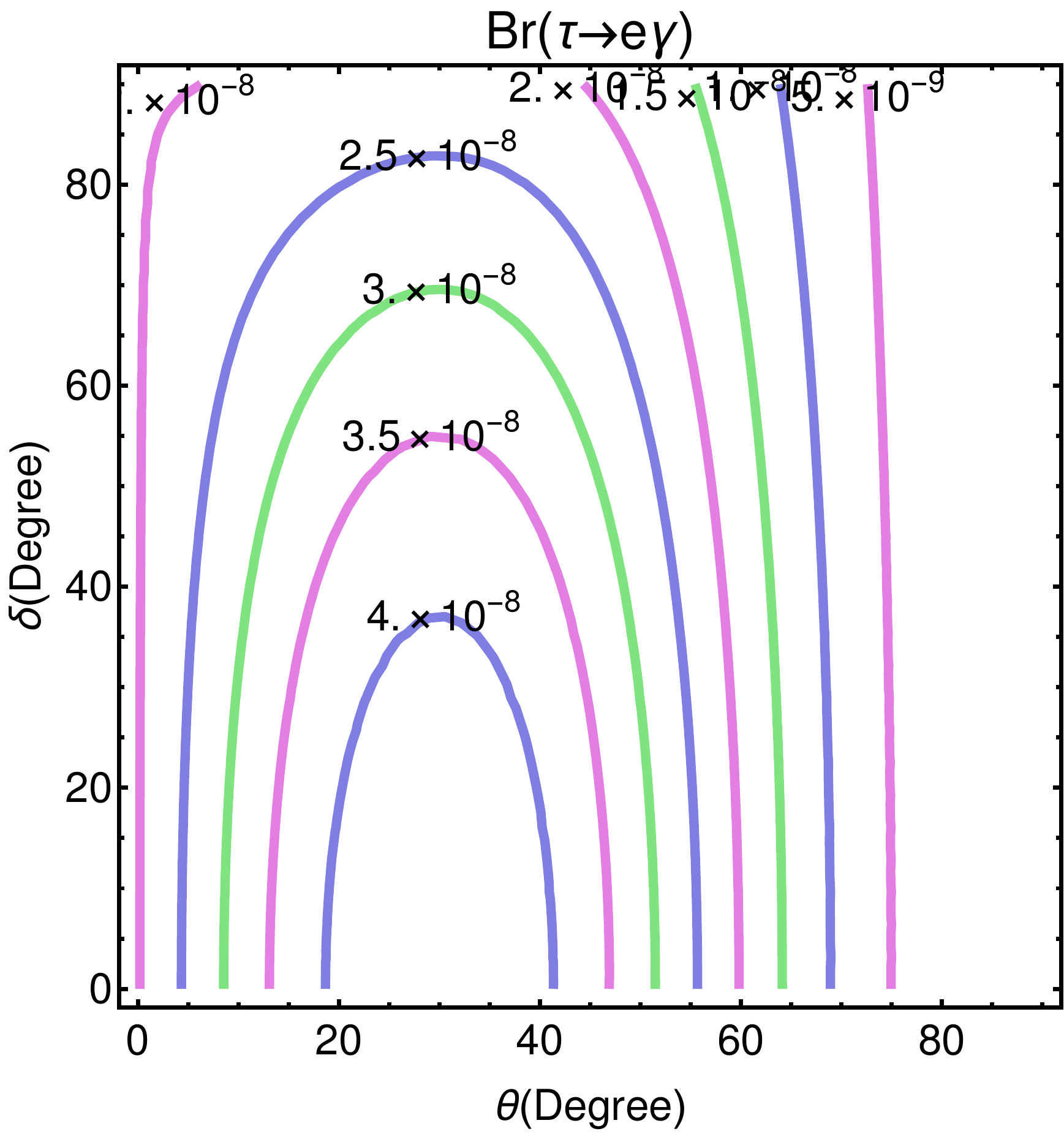} 
  \end{minipage}
\caption{Correlation plot between the internal mixing angle $\theta$ and phase $\delta$ for branching ratios for the lepton flavour 
violating processes like ${\rm BR}(\tau \to e \gamma)$ and ${\rm BR}(\tau \to \mu \gamma)$.}
\label{fig:branchingR2}
\end{figure}

Left-right symmetric model with linear seesaw mechanism can mediate other LFV processes like $Br(\mu\to eee)$ and conversion rate process $R^N(\mu\to e)$ in a 
nucleus which is discussed in reference~\cite{Cirigliano:2004mv}. The experimental bounds on these LFV processes are as follows~\cite{TheMEG:2016wtm,Nomura:2019xsb,Aubert:2009ag,Adam:2013mnn}, 
\begin{align}
\label{eq:Bexpllgamma}
	Br_{\rm exp}(\mu\to e\gamma) &< 5.7 \cdot 10^{-13}, \nonumber\\
	R^{Au}_{\rm exp}(\mu\to e)   &< 8.0 \cdot 10^{-13}, \\
	Br_{\rm exp}(\mu\to eee)     &< 1.0 \cdot 10^{-12}. \nonumber 
\end{align}
At present the process $Br(\mu\to eee)$ gives the most restrictive bound while the currently running MEG experiment~\cite{TheMEG:2016wtm,Adam:2011ch} may provide 
a better sensitivity with
\begin{equation}
\label{eq:BexpllgammaMEG}
	Br_{\rm MEG} (\mu\to e\gamma) \approx 10^{-13},
\end{equation}
Other planned experiments like COMET and Mu2e aim to reach~\cite{Kutschke:2011ux, Kurup:2011zza}
\begin{equation}
\label{eq:RmueCOMET}
	R^{Al}_{\rm COMET} (\mu\to e) \approx 10^{-16}.
\end{equation}
In Fig.\ref{fig:branchingR1} we have presented a correlation plot between the internal mixing angle $\theta$ and phase $\delta$ by fixing $M_1$,$M_2$ and $M_3$ 
at 35~GeV,100~GeV and 2~TeV respectively for branching ratios 
for the LFV process $(\mu \to e \gamma)$. The curves in the plot represent different allowed values of branching ratios for the process for different 
values of $\delta$ and $\theta$. It is seen that the values are sensitive to the current experimental bound on branching ratio as mentioned 
in eq.\ref{eq:Bexpllgamma}. Similarly in Fig.\ref{fig:branchingR2} we have shown correlation plots between $\theta$ and $\delta$ for branching ratios of the processes 
$(\tau \to e \gamma)$ and $(\tau \to \mu \gamma)$ by fixing $M_1$,$M_2$ and $M_3$ at 2.5~GeV,100~GeV and 2~TeV respectively .

\section{CP-violation for active neutrinos via Jarlskog invariants}
\label{sec:cp}
The CP-violating effects in neutrino oscillation are studied mostly in various long-baseline experiments with neutrinos 
$\nu_{\mu}$ and anti-neutrinos $\overline{\nu}_{\mu}$. This effect is characterized by the $\mbox{PMNS}$ mixing 
matrix $\mathbb{N}$ containing non-unitarity information rather than the $U_{\rm PMNS}$ matrix through Jarlskog invariant~\cite{Jarlskog:1985cw,PhysRevLett.55.1039}. 
They measure the strength of leptonic CP-violation in neutrino oscillations. 
Using usual PMNS mixing matrix $U_{\rm PMNS} \equiv U$, the standard contribution 
to these CP-violating effects is determined by the rephasing invariant 
$J_{\rm CP}$ associated with the Dirac phase $\delta_{CP}$ and matrix elements of the PMNS matrix 
$$J_{\rm CP}\equiv \text{Im}\left(U_{\alpha\, i} U_{\beta\, j} U^*_{\alpha\,j} U^*_{\beta\, j}\right) 
= \cos \theta_{12}\, \cos^2 \theta_{13}\, \cos \theta_{23}\, 
                      \sin \theta_{12}\, \sin \theta_{13}\, \sin \theta_{23}\, \sin \delta_{\rm CP}.$$ 

\begin{figure}[t!]
	\centering
	\includegraphics[width=0.45\textwidth]{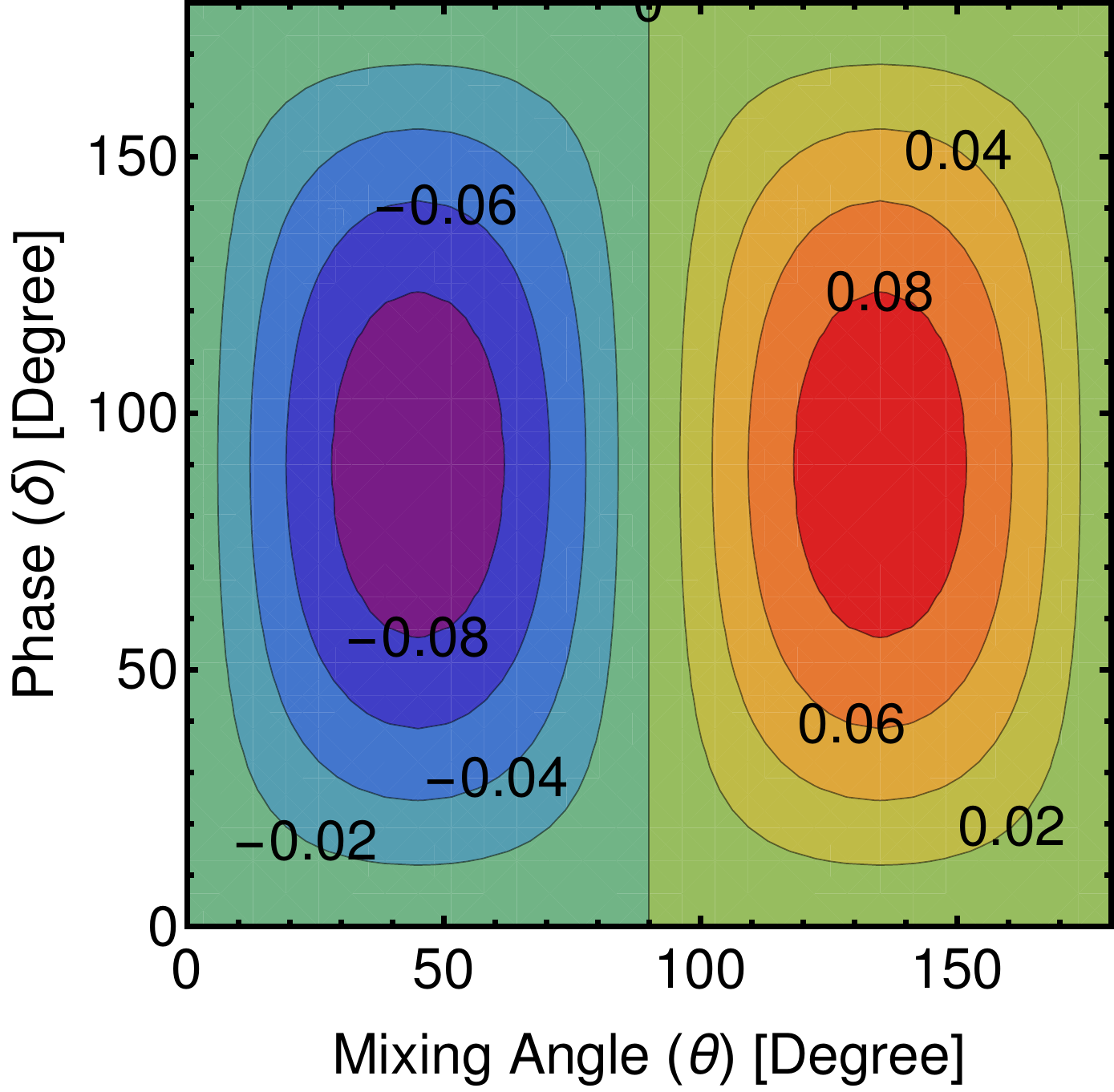}
	\caption{Plot showing relation between internal mixing angle $\theta$ and phase angle $\delta$ while fixing the rephasing invariant 
$J_{\rm CP}$ within observable range.}
\label{fig:jcp1}
\end{figure}

However, in extended seesaw schemes like linear seesaw mechanism which we follow, the leptonic CP-violation can be written 
in terms of $\mathbb{N}$ as,
\begin{eqnarray}
\mathcal{J}^{ij}_{\alpha \beta} =  \text{Im}\left(\mathbb{N}_{\alpha\, i} 
\mathbb{N}_{\beta\, j} \mathbb{N}^*_{\alpha\,j} \mathbb{N}^*_{\beta\, j}\right)
\simeq J_{\rm CP} + \Delta J^{ij}_{\alpha \beta}\, 
\end{eqnarray}
Here the indices $\alpha \neq \beta$ run over $e, \mu, \tau$ while indices $i,j$ run over $1,2,3$. 
Assuming $\sin\theta_{13}$ small and non-unitarity parameter $\eta$ (up to second order), the derived 
expression for $\Delta J^{ij}_{\alpha \beta}$ is given by~\cite{Awasthi:2013ff}
\begin{eqnarray}
\Delta J^{ij}_{\alpha \beta} = - \sum_{\gamma=e, \mu, \tau} & &\text{Im} \bigg[  
                      \eta_{\alpha \gamma}\, U_{\gamma i}\, U_{\beta j}\, U^*_{\alpha j}\, U^*_{\beta i}  
                    + \eta_{\beta \gamma}\, U_{\alpha i}\, U_{\gamma j}\, U^*_{\alpha j}\, U^*_{\beta i} 
\nonumber \\  &&     + \eta^*_{\alpha \gamma}\, U_{\alpha i}\, U_{\beta j}\, U^*_{\gamma j}\, U^*_{\beta j} 
                    + \eta^*_{\beta \gamma}\, U_{\alpha i}\, U_{\beta j}\, U^*_{\alpha j}\, U^*_{\gamma j} \bigg]\, .
\end{eqnarray}

Table.\ref{tab:JCP} shows the extra contributions to $\Delta J^{ij}_{\alpha \beta}$ due to unitary violation in the neutrino sector. 
This has been worked out by choosing different values of $M_1, M_2, M_3$ while fixing $\theta$ and $\delta$ and following the mass hierarchy 
for linear seesaw mechanism i.e, $m_{RS} >> m_{LR} >>  m_{LS}$.

\begin{table}[htb]
\centering
\begin{tabular}{|c|c|c|c|c|c|}
\hline\hline
M&$\Delta{J}^{12}_{e\mu}$& $\Delta{J}^{12}_{\mu \tau}$
&$\Delta{J}^{13}_{e\tau}$&$\Delta{J}^{23}_{e\tau}$ &$\Delta{J}^{13}_{\mu \tau }$\\ \hline
(a) &$2.77\times 10^{-6}$ &$7.2\times 10^{-5}$ &$-1.62\times 10^{-5}$ &$-2.4\times 10^{-5}$&$-7.2\times 10^{-5}$ \\ 
(b) &$2.8\times 10^{-6}$ &$5.54\times 10^{-6}$&$-5.86\times 10^{-7}$ &$-1.01\times 10^{-6}$&$-5.54\times 10^{-4}$  \\ 
%
(c) &$1.38\times 10^{-6}$ &$1.38\times 10^{-6}$ &$1.38\times 10^{-6}$&$-1.38\times 10^{-6}$&$-1.38\times 10^{-6}$\\ \hline
\hline
\end{tabular}
\caption{Estimated CP-violating effects for three cases, (a) $M=(10,50,1500)$~GeV, (b) partially degenerate masses  
$M=(50, 50, 1000)$~GeV and (c) fully degenerate masses  
$M=(500, 500, 500)$~GeV. We have fixed the internal mixing angle and phase $\delta$ as $120$~degree and $60$~degree, respectively. }
\label{tab:JCP}
\end{table}

\begin{figure}[t!]
	\centering
    \includegraphics[width=0.45\textwidth]{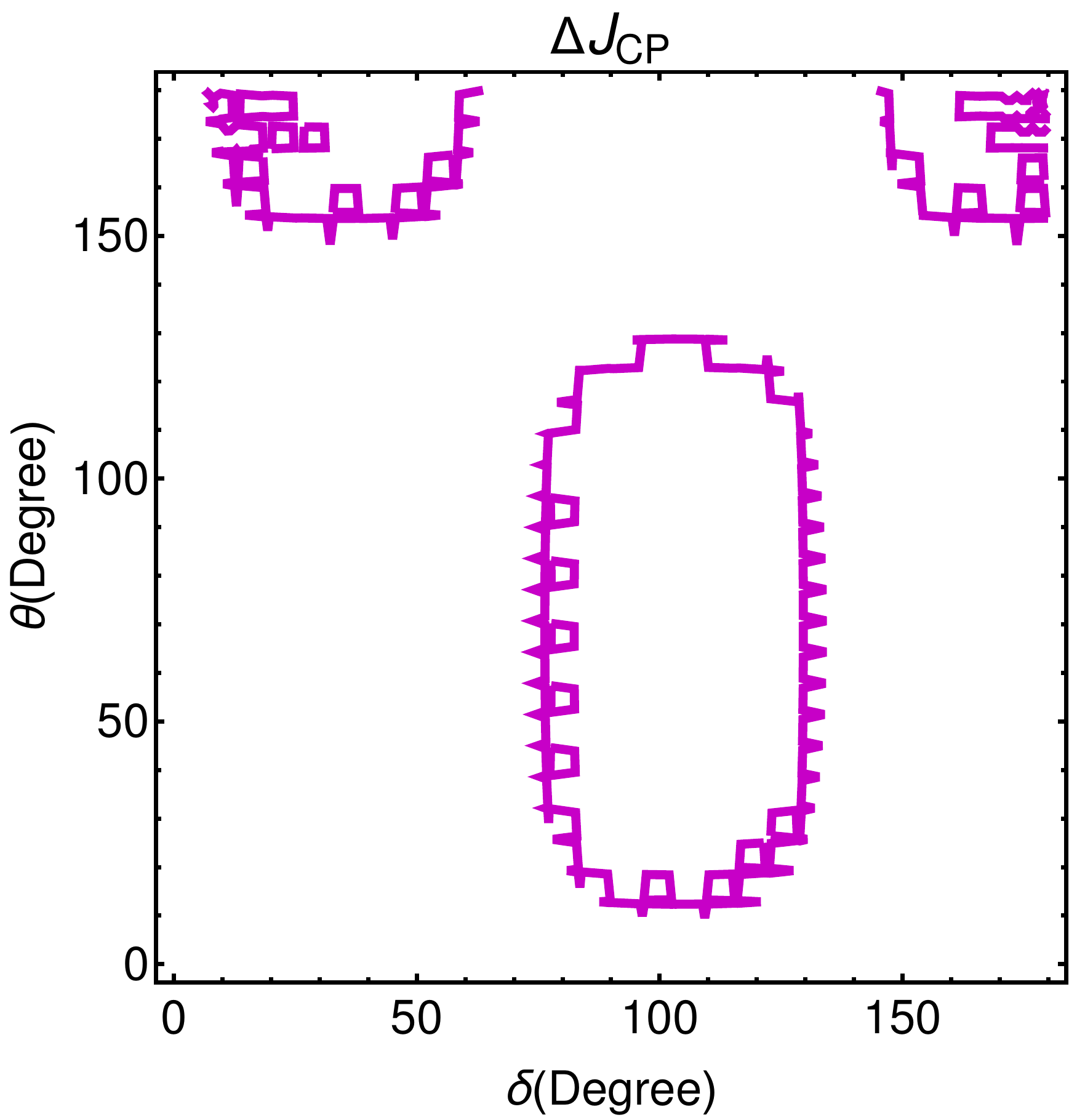}
	\includegraphics[width=0.45\textwidth]{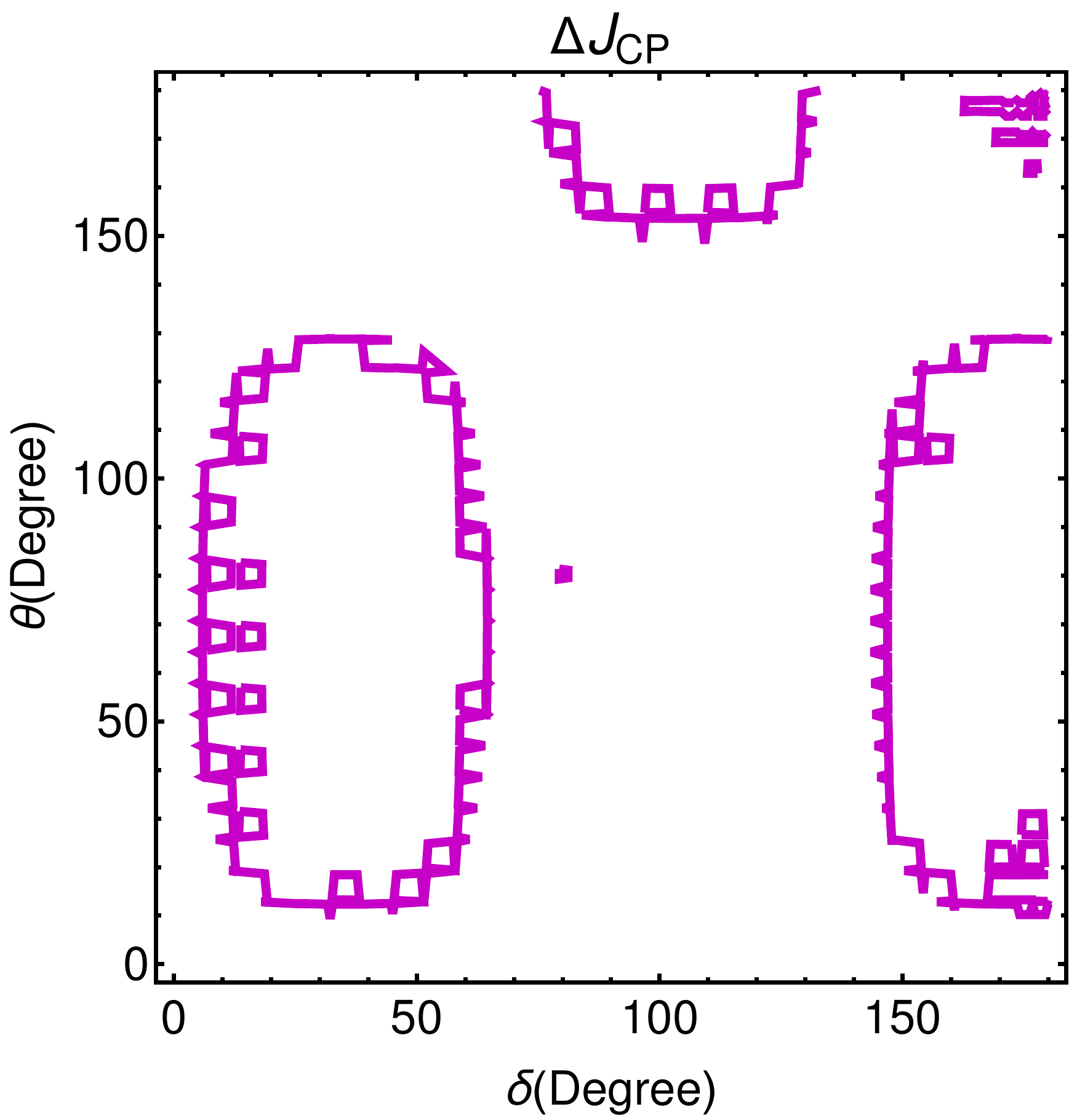}
	\caption{The contour plot for the rephasing inavariant $\Delta J^{12}_{e \mu}$ due to unitary violation in the neutrino sector in the plane of 
    internal mixing angle $\theta$ and internal phase angle $\delta$. The value of $\Delta J^{12}_{e \mu}$ are considered here around $10^{-6}$ by taking degenerate values of $M_{1}, M_2, M_3$.}
	\label{fig:jcp2}
\end{figure}

The numerical results for relation between the internal mixing angle $\theta$ and the phase $\delta$ are displayed in Fig.\ref{fig:jcp1} and \ref{fig:jcp2}.
In Fig.\ref{fig:jcp1} the range for the value of $J_{\rm CP}$ comes out to be 0.02 to 0.04 which matches with the experimental observable range. 
Since $J_{\rm CP}$ has a modulus, both negative and positive values are shown in the figure. In Fig.\ref{fig:jcp2} 
the value of $\Delta J^{12}_{e \mu}$ lies around $10^{-6}$ for degenerate values of $M_1, M_2, M_3$.
The allowed range of rephasing invariant $J_{\rm CP}$ is $0.026 < |J_{\rm CP} | < 0.036$ and that of $\Delta J_{\rm CP}$ is  $10^{-5} < |\Delta J_{\rm CP} | < 10^{-7}$ .

\section{Conclusion}
\label{sec:conclusion}
We have studied a left-right symmetric model with discrete $A_4$-flavour symmetry where neutrino masses and mixing are explained via linear 
seesaw mechanism. Even though $A_4$ flavour based models have been studied before, we have shown here that the $A_4$ extension of LRSM simpler 
analytical expressions for large non-unitarity effect which can lead to dominant contributions to LFV decays.
We have shown correlation among model parameters like internal mixing angle $\theta$, internal phase $\delta$ and their dependence 
on experimentally determined parameters like mixing angles $\theta_{12}$, $\theta_{23}$, $\theta_{13}$ and sum of neutrino masses both 
analytically as well as numerically.

The model facilitates sizable charged lepton flavour violation due to contributions from heavy right handed neutrinos. 
We have studied non-unitarity effects in linear seesaw which has implications on prediction for LFV decays 
like $\mu\to e\gamma$, $\mu\to eee$ and $\mu\to e$ and by saturating the experimental bounds on these decays 
we have derived constraints on input model parameters. Finally we have studied CP-violation for active neutrinos via Jarlskog invariants 
and shown extra contributions to CP violating effects that the model generates due to unitarity violation in the neutrino sector. 
Again by saturating the experimental values of unitarity violating parameter $\eta$ in $e\mu, e\tau, \mu\tau, \tau\tau$ sectors 
we have generated contour plots to show constraints on model parameters. Interestingly, the range for the value of $J_{\rm CP}$ comes out 
to be 0.02 to 0.04 which matches with the experimental observable range. 

\section{Acknowledgement}
Purushottam Sahu would like to acknowledge Ministry of Human Resource Development (MHRD), Govt of India for financial support. 

\bibliographystyle{utcaps_mod}
\bibliography{A4}
\end{document}